\documentclass[prd,twocolumn,showpacs,superscriptaddress,footinbib]{revtex4-1}
\usepackage{amsfonts}
\usepackage{amsmath}
\usepackage{amssymb}
\usepackage{bm}
\usepackage{dcolumn}
\usepackage{graphicx}
\usepackage{graphics}
\usepackage{latexsym}
\usepackage{rotating}
\usepackage[colorlinks=true]{hyperref}
\usepackage[all]{hypcap} 
\usepackage{xspace} 
\usepackage[usenames]{xcolor}
\usepackage{mathrsfs}
\usepackage{multirow}
\usepackage{enumitem}

\usepackage{ulem}
\definecolor {darkgreen}{rgb}{0.2,0.7,0.2}

\newcommand\be{\begin{equation}}
\newcommand\ee{\end{equation}}
\newcommand\bw{\begin{widetext}}
\newcommand\ew{\end{widetext}}

\newcommand{\bea}{\begin{eqnarray}}
\newcommand{\eea}{\end{eqnarray}}

\usepackage{cellspace} %
\def\apj{Astrophys. J. \ }%
\def\apjl{Astrophys. J. Lett. \ }%
\def\mnras{MNRAS}%
\def\prd{Phys. Rev. D \ }%
\begin{document}
\title{Revisiting the maximum mass of differentially rotating neutron
  stars\\ in general relativity: \"Ubermassive stars with realistic
  equations of state}

\author{Pedro L. Espino}
\affiliation{Department of Physics, University of Arizona, Tucson, AZ 85721, USA}
\author{Vasileios Paschalidis}
\affiliation{Department of Physics, University of Arizona, Tucson, AZ 85721, USA}
\affiliation{Department of Astronomy, University of Arizona, Tucson, AZ 85721,USA}

\begin{abstract}
We study the solution space of general relativistic, axisymmetric,
equilibria of differentially rotating neutron stars with realistic,
nuclear equations of state. We find that different types of stars,
which were identified by earlier works for polytropic equations of
state, arise for realistic equations of state, too. Scanning the
solution space for the sample of realistic equations of state we
treat, we find lower limits on the maximum rest masses supported by
cold, differentially rotating stars for each type of stars. We often
discover equilibrium configurations that can support more than 2 times
the mass of a static star. We call these equilibria ``\"ubermassive'',
and in our survey we find \"ubermassive stars that can support up to
2.5 times the maximum rest mass that can be supported by a cold,
non-rotating star with the same equation of state. This is nearly two
times larger than what previous studies employing realistic equations
of state had found, and which did not uncover \"ubermassive neutron
stars. Moreover, we find that the increase in the maximum rest mass
with respect to the non-spinning stellar counterpart is larger for
moderately stiff equations of state. These results may have
implications for the lifetime and the gravitational wave and
electromagnetic counterparts of hypermassive neutron stars formed
following binary neutron star mergers.
\end{abstract}

\date{\today}
\maketitle


\section{Introduction}

Hypermassive Neutron Stars (HMNSs)~\cite{HypermassiveNSorig} are
transient configurations that are supported against gravitational
collapse by the additional centrifugal support provided by
differential rotation, and possibly also by thermal
pressure~\cite{Sekiguchi:2011zd,Paschalidis:2012ff}.  HMNSs may be
ubiquitous remnants of binary neutron star (BNS) mergers (see
e.g.~\cite{Faber:2012rw,Paschalidis:2016agf,Baiotti:2016qnr,Paschalidis:2016vmz}
for reviews and references therein). An HMNS was also a likely
outcome~\cite{Shibata:2017xdx,Margalit2017,Ruiz:2017due,Zhang:2017fsy}
 of the LIGO/Virgo event
GW170817~\cite{TheLIGOScientific:2017qsa,multimessGW170817}.

The study of differentially rotating relativistic stars is useful for
understanding the types of BNS merger remnants that are possible, and
their properties. Modest to high degrees of differential rotation may
support an HMNS against collapse on dynamical timescales, but such
objects are unstable on secular timescales (see
e.g.~\cite{dlsss06a,dlsss06b,Paschalidis:2012ff} and references
therein). An important quantity that determines whether following a
BNS merger there will be prompt, delayed or no collapse at all, is the
maximum mass that can be supported given an equation of
state. Studying general relativistic, equilibrium models of
differentially rotating stars provides a straightforward approach to
determine this maximum mass.

In~\cite{Lyford:2002ip,Morrison:2004fp} it was shown that cold,
axisymmetric, differentially rotating stars described by either
polytropic or realistic equations of state (EOSs) can support up to approximately $70 \%$ more mass when compared to the maximum rest mass that can be
supported by a non-rotating star -- the Tolman-Oppenheimer-Volkoff
(TOV) limit. This result holds for the differential rotation law
of~\cite{KEH1989MNRAS.237..355K} which we refer to as the KEH
law. However, in~\cite{Ansorg:2009} it was pointed out that early
efforts to find the maximum rest mass of differentially rotating,
axisymmetric configurations did not account for the full solution space with the
KEH law. Subsequently, it was found in~\cite{Gondek-Rosinska:2016tzy} that
differentially rotating, axisymmetric, $\Gamma=2$ polytropic models of
neutron stars built with the KEH law can support up to $\sim 4$ times
the TOV limit at even modest degrees of differential rotation.

The solution space for relativistic, differentially rotating,
axisymmetric stars with the KEH law has been shown to exhibit four
types of equilibrium solutions~\cite{Ansorg:2009} labeled A, B, C, and
D. A careful scan among these types reveals that stars with
quasi-toroidal topology are those that tend to be the most
massive. Each stellar configuration belonging to a solution type falls
along a sequence characterized by a quadruplet of parameters
consisting of the maximum energy density $\epsilon_{\rm max}$, the
degree of differential rotation $\hat{A}^{-1}$, the ratio of polar to
equatorial radius $r_p/r_e$, and the parameter $\hat \beta$ describing
how close to the mass-shedding limit the configuration is. Note that
the first three of the above parameters are needed to completely
specify a configuration, yet the solution space requires four
parameters to be described. The full solution space with the KEH law
has been studied in great detail for polytropic EOSs of varying
stiffness~\cite{Studzinska:2016ofb}. In~\cite{Studzinska:2016ofb} it
was further shown that the existence of four types of solutions is a
Universal feature for a range of polytropic indices $n \in
[2/3,2]$. Nevertheless, for $n=1.5$ the authors did not report stars
of type B, C, or D. These results imply that the possible types of
solutions may depend on the equation of state. This is important
because neutron star EOSs are not described by a single polytropic
index and different realistic nuclear equations of state have varying
degrees of stifness. While $n=1.5$ does not correspond to models of
neutron stars, a natural question arises by the work
of~\cite{Gondek-Rosinska:2016tzy,Studzinska:2016ofb}: do the different
types of differentially rotating, axisymmetric stars arise for
realistic nuclear EOSs? If they do arise, what is the maximum rest
mass that can be be supported by the different types of solutions when
realistic nuclear EOSs are considered?

In this paper, we address these questions by considering the solution
space for differentially rotating, axisymmetric stars built with the
KEH law with realistic nuclear EOSs. We find that the different types
of solutions identified in~\cite{Ansorg:2009} arise even for realistic
neutron star matter. As
in~\cite{Gondek-Rosinska:2016tzy,Studzinska:2016ofb} we find that many
configurations can support a mass more than 2 times the TOV limit. We
term these configurations ``\"ubermassive''.  We propose a different
name for these because \"ubermassive neutron stars (\"UMNS) cannot
arise in Nature through mergers of binary neutron stars. Thus, if
\"UMMNs ever form through astrophysical processes, it would have to be
through some other more exotic channels. For the sample of realistic
EOSs we explore, in our scan of the solution space we find
\"UMNSs that can support up to 2.5 times (150 \% more mass
than) the corresponding TOV limit.  Moreover, the increase in the
maximum rest mass with respect to the TOV limit is larger for
moderately stiff equations of state.

The remainder of this paper is organized as follows.  In Section
~\ref{sec:review} we review basic equations and details pertaining to
the solution space of differentially rotating stars built with the KEH
law. In Section ~\ref{sec:EOSs} we present the EOSs we treat here and
their basic properties. In Section ~\ref{sec:methods} we describe our
methods and reproduce some of the results presented
in~\cite{Gondek-Rosinska:2016tzy} for a $\Gamma=2$
polytrope. Section~\ref{sec:results} details our results, showing the
solution space of differentially rotating stars with realistic
nuclear EOSs along with the maximum rest mass models we found for each EOS
we considered. We conclude in Section ~\ref{sec:conclusions} with a
summary of our findings and a discussion of future
directions. Geometrized units, where $G=c=1$, are adopted throughout,
unless otherwise specified.

\section{Basic equations and types of differentially rotating stars}
\label{sec:review}

The spacetime of stationary, axisymmetric, equilibrium rotating stars
is described by the following line element in spherical polar coordinates (see e.g.~\cite{CST94a})
\begin{equation}
  \label{eq:ds}
ds^2 = -e^{\gamma+\rho}dt^2+e^{2\alpha}(dr^2+r^2d\theta^2)+e^{\gamma-\rho}r^2\sin^2\theta(d\phi-\omega dt)^2,
\end{equation}
where the metric potentials $\gamma, \rho, \alpha$ and $\omega$ are
functions of $r$ and $\theta$ only, and are determined by the solution
of the Einstein equations coupled to the hydrostationary equilibrium
equation for perfect fluids (see e.g.~\cite{Paschalidis:2016vmz} for a
review and other forms of the line element used in the literature). To
close the system of equations an EOS and a differential rotation law are required.

Most studies of differentially rotating stars adopt the KEH rotation
law~\cite{KEH1989MNRAS.237..355K}, which is also called j-constant
rotation law (see~\cite{Paschalidis:2016vmz} for a summary of other
differential rotation laws.). In this law the specific angular
momentum is a function of the angular velocity as follows
\begin{equation} \label{eq:Komatsurotlaw}
u^tu_\phi = A^2(\Omega_c - \Omega),
\end{equation}
where $u^t$ and $u^\phi$ are the temporal and azimuthal components of
the fluid four velocity, respectively, $\Omega = u^\phi/ u^t $ is the
local angular velocity of the fluid as seen by an observer at
infinity, and $\Omega_c$ is the angular velocity on the rotation axis. It
is common and convenient to parameterize the angular velocity by
considering the ratio of polar ($r_p$) to equatorial ($r_e$) radius of
the star, $\frac{r_p}{r_e}$. Stars with larger values of $\Omega_c$
tend to have a smaller value of $\frac{r_p}{r_e}$, indicative of a
``flatter'' stellar shape. The parameter $A$ in
Eq.~\eqref{eq:Komatsurotlaw} has units of length and is a measure of
the degree of differential rotation in the star, i.e, the lengthscale over which the fluid angular velocity changes in the star. It is also common to
use a rescaled $A$ parameter
\begin{equation} \label{eq:defAm1}
\hat{A}^{-1} = \dfrac{r_e}{A}.
\end{equation}
A general relativistic stellar configuration is then completely
determined by the values of $\hat{A}^{-1}$, $\frac{r_p}{r_e}$, and the
central or maximum energy density ($\epsilon_{\rm max}$). In the case
of uniform rotation or cases with low degrees of differential rotation
the central energy density and $\epsilon_{\rm max}$ coincide, since
$\epsilon_{\rm max}$ occurs at the center of the star. However, when
considering differentially rotating stars a quasi-toroidal topology
may arise in which case $\epsilon_{\rm max}$ is not at the geometric
center of the configuration. In these cases it is more convenient to
specify $\epsilon_{\rm max}$ instead of the value of the energy
density at the center of the star. Models with extreme quasi-toroidal shapes
tend to have very small (but non-zero) densities near the center.

The parameter $\hat{A}^{-1}$ is important for identifying the
different types of solutions that arise for rotating stars. When
$\hat{A}^{-1}=0$ the stars are uniformly rotating, while stars with
$\hat{A}^{-1}\neq 0$ are differentially rotating.  Models with relatively high
values of $\hat{A}^{-1}$ (typically $\hat{A}^{-1} \gtrsim 1.0)$ tend
to show a smooth transition from spheroidal to quasi-toroidal
topologies, depending on the values of $\frac{r_p}{r_e}$ and
$\epsilon_{\rm max}$. Models with lower values of $\hat{A}^{-1}$
(typically $\hat{A}^{-1} \lesssim 0.7$ for the values of
$\epsilon_{\rm max}$ considered here) show a richer solution
space, as we discuss below.

Another important parameter in describing differentially rotating
stars is $\beta$, which parametrizes how close to mass-shedding the
stellar model is. The parameter $\beta$ was introduced in
\cite{Ansorg:2009} and is defined as
\begin{equation}\label{eq:defbeta}
\beta = -\left(\dfrac{r_e}{r_p}\right)^2 \dfrac{d(z^2)}{d(\varpi^2)} \biggr\rvert_{\rho=r_e},
\end{equation}
where $\varpi=r\sin(\theta)$ and $z=r\cos(\theta)$ are cylindrical
coordinates, and the derivative is evaluated on the surface of the
star at the equator. On the surface of the star $r=r(\theta)$, thus the
function $z^2(\varpi^2)$ describes the surface shape, whose slope at
the stellar equator determines how close to mass-shedding the
configuration is. The ``mass-shedding parameter" is defined in
terms of $\beta$ as~\cite{Ansorg:2009}
\begin{equation}\label{eq:defbetahat}
\hat{\beta} = \dfrac{\beta}{1 + \beta}.
\end{equation}

While $\hat{\beta}$ is not a gauge-invariant quantity, it is useful in
describing models in coordinates such as those defined by
Eq. \ref{eq:ds}. Depending on the surface slope at the equator,
$\hat{\beta}$ will approach different values. We are generally
interested in three limiting values of $\hat{\beta}$:

\begin{enumerate}[leftmargin=*,noitemsep]
  
\item Non-rotating, spherical limit: For a spherical TOV star,
  $\frac{r_p}{r_e} = 1$, and the derivative
  $\dfrac{d(z^2)}{d(\varpi^2)} = -1$ everywhere on the surface. Thus,
  in this limit $\hat{\beta} \longrightarrow \frac{1}{2}$.\\

\item Mass-shedding limit: At the mass-shedding limit, the stellar
  configuration begins to lose mass at the equator. The surface
  derivative at the equator vanishes $\dfrac{d(z^2)}{d(\varpi^2)} =
  0$.  Hence, $\hat{\beta} \longrightarrow 0 $ at the mass-shedding
  limit.\\

\item Toroidal limit: As the stellar topology approaches that of a
  toroid, $r_p \longrightarrow 0$, and ${\beta}$ becomes large. This
  implies that $\hat{\beta}\longrightarrow 1$ as a sequence approaches
  the toroidal limit.

\end{enumerate}

The above discussion suggests that the complete set of parameters
describing general relativistic equilibria of stationary and
axisymmetric, differentially rotating stars with the KEH law is the
quadruplet $(\epsilon_{\rm max}, \frac{r_p}{r_e}, \hat{A}^{-1},
\hat{\beta})$.

The solution types can be distinguished by specifying $\epsilon_{\rm
  max}$ and considering the limiting values of $\hat{\beta}$ for
sequences of constant $\hat{A}^{-1}$ in the $(\frac{r_p}{r_e},
\hat{\beta})$ plane. This requires that one slowly vary the quadruplet
$(\epsilon_{\rm max}, \frac{r_p}{r_e}, \hat{A}^{-1}, \hat{\beta})$ to
carefully scan the space of solutions. We use the convention
introduced in \cite{Ansorg:2009} to distinguish the types of
differentially rotating stars at fixed $\epsilon_{\rm max}$ for
sequences of constant $\hat{A}^{-1}$. Given that in the numerical
construction of rotating stars we always start with an initial guess
solution corresponding to a static star, and then slowly vary the
stellar parameters to reach a particular type of solution at fixed
$\epsilon_{\rm max}$, below we list the general trajectory of
solutions used in building the corresponding sequences:\\

\begin{itemize}[leftmargin=*,noitemsep]

\item Type A: This sequence of solutions consists strictly of
  spheroids.  For low degrees of differential rotation (i.e, close to
  rigid rotation), stars are spheroidal. Spinning these stars up (i.e,
  decreasing $\frac{r_p}{r_e}$) results in mass-shedding, so that the
  Type A sequence goes from the limiting solution of spherical stars
($\frac{r_p}{r_e} =1, \hat{\beta} = 0.5$) to mass-shedding 
($\hat{\beta}=0$). Starting from a spherical
  solution, these models are obtained by simply spinning up the
  initial model. A potential path in the parameter space is as follows   \\
  
Spheroid (low $\hat{A}^{-1}$) $\longrightarrow$ decrease $\dfrac{r_p}{r_e}$ $\longrightarrow$ Mass-shedding\\

\item Type B: This type of star often exits for the same values of
  $\hat{A}^{-1}$ as Type A stars, but at lower values of
  $\frac{r_p}{r_e}$. Spinning these stars down (increasing
  $\frac{r_p}{r_e}$) results eventually in mass-shedding. Therefore,
  the Type B sequence goes from the limiting solution of toroids
  $(\hat{\beta} \longrightarrow 1.0)$ to mass-shedding
  $(\hat{\beta}=0)$.
  These models can be reached numerically by spinning up an initial spherical model
  (decreasing $\frac{r_p}{r_e}$) with high $\hat A^{-1}$ to obtain
  quasi-toroidal solutions, then decreasing $\hat A^{-1}$, and finally
  increasing $\frac{r_p}{r_e}$ to approach the mass-shedding limit.
  A potential path in the parameter space is as follows \\

Spheroid (low $\hat{A}^{-1}$) $\longrightarrow$ increase $\hat{A}^{-1}$, decrease
$\dfrac{r_p}{r_e}$ $\longrightarrow$ Quasi-toroid (high $\hat{A}^{-1}$) $\longrightarrow$ 
decrease $\hat{A}^{-1}$ $\longrightarrow$ Quasi-toroid (low $\hat{A}^{-1}$) $\longrightarrow$
increase $\dfrac{r_p}{r_e}$ $\longrightarrow$ Mass-shedding.\\

The Type B stars near the mass-shedding limit are difficult to reach and we were not able to construct such extreme configurations.\\

\item Type C:
	This sequence exhibits a smooth transition from a 
    spherical solution $(\hat{\beta} = 0.5)$ to a quasi-toroidal solution 
    $(\hat{\beta} = 1.0)$. 
    As such, starting at a 
    spheroid with high $\hat A^{-1}$, and spinning Type C stars up 
    by decreasing $\frac{r_p}{r_e}$ would not result in mass-shedding, 
    but would shape the models into a quasi-toroid. A potential path in the parameter space is as follows \\
    
  Spheroid (low $\hat{A}^{-1}$) $\longrightarrow$ increase $\hat{A}^{-1}$, decrease
$\dfrac{r_p}{r_e}$ $\longrightarrow$ Quasi-toroid (high $\hat{A}^{-1}$)\\

\item Type D: This type typically covers the smallest part of the parameter
  space. The models of this type are non-trivial to build directly
  from a spherical solution. This is because Type D sequences start and
  end at the mass-shedding limit ($\hat{\beta}=0$). Spinning these
  stars either up or down would result in mass-shedding. We were
  unable to build Type D sequences at fixed values of
  $\hat{A}^{-1}$ for any of the cases considered. However, we were
  able to construct individual \textit{candidate} Type D models at specific values of the
  quadruplet $(\epsilon_{\rm max}, \frac{r_p}{r_e}, \hat{A}^{-1},
  \hat{\beta})$.

\end{itemize}

\section{Equations of State}
\label{sec:EOSs}

We consider a set of four realistic EOSs, all of which can be found on
the Compstar Online Supernovae Equations of State
(ComPOSE)~\cite{Typel:2013rza} database. We chose two zero-temperature
EOSs and two finite temperature EOSs (in their ``cold'' limit) to
study. The zero-temperature, nuclear EOSs we considered are
APR~\cite{PhysRevC.58.1804} and FPS~\cite{Lorenz:1992zz}. 
These zero-temperature EOSs were also considered in~\cite{Morrison:2004fp}
and were chosen for a suitable comparison to the maximum rest mass models
found therein.

\begin{figure}[htb]
  \includegraphics[width=8.5cm]{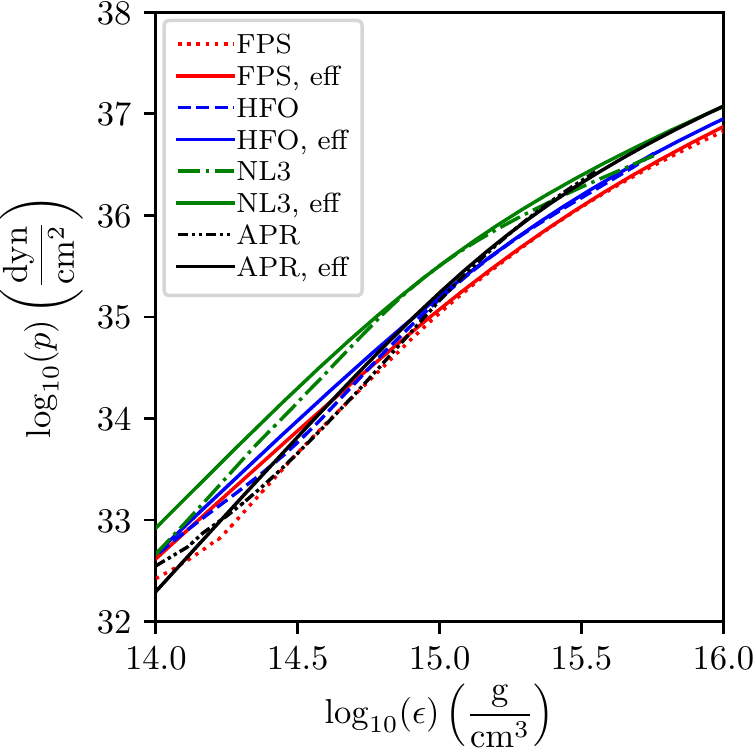}
  \caption{Pressure as a function of energy density for the EOSs in our
    sample. The red dashed, blue dotted, green dash-dotted, and black
    dash-double-dotted lines correspond to the FPS, HFO, NL3, and APR
    EOSs, respectively. The solid lines of the same color scheme
    correspond to representations of each EOS using a single polytrope
    as described by Equations ~\eqref{eq:polyeff_p}
    and~\eqref{eq:polyeff_e}.}
  \label{fig:EOS}
\end{figure}

The first finite temperature EOS we consider is a variant of the EOS
of~\cite{Shen:2011kr} which includes electrons, protons, neutrons and
will hereafter be referred to as NL3. We also consider the EOS
of~\cite{Hempel:2009mc}, which will hereafter be referred to as
HFO. The finite temperature EOS tables include values of the rest mass
density $\rho_0$ at different values of the temperature $T$ and the
electron fraction $Y_e$. Since our focus is on cold, equilibrium
models of differentially rotating stars we set $T = 0.01$ MeV, and
enforce neutrinoless beta equilibrium as is common in the case of
finite temperature EOSs. In particular we numerically solve for the
value of $Y_e$ at which chemical equilibrium is established between
neutrons, protons, and electrons,
\begin{equation}\label{eq:betaequil}
\mu_n - \mu_p - \mu_e = 0,
\end{equation}
where $\mu_i$ is the chemical potential of species $i$. Once $\rho_0$
and $T$ are specified for the EOS tables, we scan through values of
$Y_e$ until the condition in Equation (\ref{eq:betaequil}) is met. We
then change the value of $\rho_0$ and repeat, building a tabulated EOS
of pressure, rest mass density and energy density for the set of electron
fractions corresponding to beta equilibrium. Figure~\ref{fig:EOS}
shows a plot of pressure as a function of energy density for the set of EOSs we
treat in this work. Among these EOSs the FPS one does not satisfy the
$2M_\odot$ lower bound on the maximum rest mass of a TOV
star~\cite{Demorest2010,Antoniadis2013}, but we include it in our set
to compare with~\cite{Morrison:2004fp}. Also, the NL3 EOS may have too
high a maximum
mass~\cite{Margalit2017,Ruiz:2017due,Rezzolla:2017aly,Shibata:2017xdx}
and a radius of a $1.4M_\odot$ star that may be large compared to
constraints from the observation of
GW170817~\cite{TheLIGOScientific:2017qsa,Abbott:2018exr,Raithel:2018ncd},
but we include it to examine the increase in the maximum rest mass due to
differential rotation for an EOS with a large TOV limit mass.

\begin{table}[htb]
 	\centering
 	\caption{Ratio of average energy density to maximum energy
          density $C_\epsilon^{1.4}$ (for models of rest mass $M_0 = 1.4
          M_\odot$) and effective adiabatic index
          $\Gamma_{eff}^{nuc}$ as measures of EOS stiffness for each
          EOS in our study. $M_{0,\rm max}^{TOV}$, $M_{0, \rm max}^{sup}$ are the
          rest masses of the TOV limit and the supramassive limit,
          respectively, and $M_{\rm ADM,max}^{TOV}$, $M_{\rm ADM,max}^{sup}$
          are the gravitational masses of the TOV limit and the
          supramassive limit, respectively. All masses are in units of
          $M_\odot$. \label{tab:stiffness}}
\begin{tabular}{c c c c c c c }\hline \hline 
EOS & $C_\epsilon^{1.4}$ & $\Gamma_{eff}^{nuc}$ & $M_{0, \rm max}^{TOV}$ & $M_{0, \rm max}^{sup}$ & $M_{\rm ADM,max}^{TOV}$ & $M_{\rm ADM,max}^{sup}$\\ \hline
FPS & 0.40 & 2.55 & 2.10 & 2.45 & 1.80 & 2.12\\
HFO & 0.42 & 2.66 & 2.41 & 2.83 & 2.06 & 2.44\\
NL3 & 0.43 & 2.84 & 3.27 & 3.88 & 2.75 & 3.30\\ 
APR & 0.44 & 3.07 & 2.66 & 3.09 & 2.19 & 2.60\\ \hline
\end{tabular}
\end{table}
 
We compare the EOSs in terms of their stiffness, which we characterise
by the ratio of average energy density $\bar{\epsilon}$ to maximum
energy density $\epsilon_{\rm max}$ in models of equal rest mass $M_0$
for each EOS. The average density is defined as~\cite{Morrison:2004fp}
\begin{equation}\label{eq:defe_av}
\bar{\epsilon} \equiv \dfrac{3M}{4\pi R_c^3},
\end{equation}
where $M$ is the gravitational mass and $R_c$ the circumferential
radius. We build $M_0 = 1.4 M_\odot$ TOV models for each EOS, and
look at the ratio of average to maximum energy density $C_\epsilon$
\begin{equation}\label{eq:Ce}
C_\epsilon = \dfrac{\bar{\epsilon}}{\epsilon_{\rm max}}.
\end{equation}
A maximally stiff EOS would have $C_\epsilon = 1$, corresponding to a
uniform energy density configuration. We list $C^{1.4}_\epsilon$ (Equation ~\eqref{eq:Ce} for a $1.4M_\odot$ star) for each
EOS in Table~\ref{tab:stiffness}.

As an alternative measure of EOS stiffness, and to compare with the
polytropic models in~\cite{Gondek-Rosinska:2016tzy}, we also consider
the effective adiabatic index $\Gamma_{eff}^{nuc}$ for each model,
calculated as in~\cite{Morrison:2004fp}. In particular, to find
$\Gamma_{eff}^{nuc}$ for each realistic EOS we first calculate
$C_\epsilon^{nuc}$ for the maximum rest mass TOV model. Next, we
calculate the ratio $C_\epsilon^{poly}$ for the maximum rest mass TOV
models of polytropes with a wide range of adiabatic indices
$\Gamma^{poly}$, and construct a function
$\Gamma^{poly}(C_\epsilon^{poly})$ that we interpolate at values of
$C_\epsilon^{poly}$ that are not in our table. The effective adiabatic
index of a nuclear EOS is then defined through
\begin{equation}\label{eq:Gamma_nuc}
\Gamma_{eff}^{nuc} = \Gamma^{poly}(C_\epsilon^{nuc}).
\end{equation}

The effective adiabatic indices for the EOSs we treat are listed in
Tab.~\ref{tab:stiffness}, where we also show the TOV limit mass and
the supramassive limit mass (the maximum mass that can be supported
when allowing for maximal uniform rotation) for each of these EOSs.
Compared to the values of $\Gamma_{eff}^{nuc}$ reported in
\cite{Morrison:2004fp} for the FPS and APR EOSs, our results differ by
$0.04\%$ and $1.56\%$, respectively. Note that we have ranked the EOSs in
Tab.~\ref{tab:stiffness} in order of increasing $C^{1.4}_\epsilon$ and $\Gamma^{nuc}_{eff}$. By both
metrics of the stiffness APR is the stiffest, and FPS is the softest. Using $C^{1.4}_\epsilon$ as a measure of
EOS stiffness is better suited for realistic EOSs because in using
$\Gamma_{eff}^{nuc}$ we are approximating the EOS as a single
polytrope which would not account for microphysical phenomena that can soften or stiffen the EOS at varying energy
densities. However, using $\Gamma_{eff}^{nuc}$ to measure EOS
stiffness is useful when comparing features in the solution space of
realistic EOSs to those of polytropes. All of the EOSs in our set have
$2.5 < \Gamma_{eff}^{nuc} < 3.1$, and it turns out that certain features of the
solution space for these realistic EOSs are consistent with the
$\Gamma \geq 2.5$ polytropes~\cite{Studzinska:2016ofb}, as further
discussed in Sec.~\ref{sec:results}.

In order to see how well approximated the realistic EOSs are by single
polytropes, we also include a polytropic representation of each
nuclear EOS. Along with the effective adiabatic index
$\Gamma_{eff}^{nuc}$, we calculate an effective polytropic constant
$\kappa_{eff}^{nuc}$ for each EOS as detailed in Appendix \ref{sec:appendixA}.

The polytropic representations of the nuclear EOSs are presented in Fig. ~\ref{fig:EOS}. Although not perfect, using 
a single polytrope to represent the nuclear EOS is reasonable at higher densities, 
and the qualitative results of~\cite{Studzinska:2016ofb} for polytropes 
of varying polytropic indices may be suitably compared to those presented 
in this work for nuclear EOSs.

\section{Methods}
\label{sec:methods}

We adopt the code detailed in~\cite{CST94a} and~\cite{CST94b}
(hereafter referred to as the Cook code) to solve the coupled
Einstein-hydrostationary equilibrium equations in axisymmetry. This
code was also used in~\cite{Lyford:2002ip,Morrison:2004fp}. In this
section we describe the numerical grid and tests we performed to
validate the code in the case of differentially rotating stars 
found by~\cite{Ansorg:2009,Gondek-Rosinska:2016tzy,Studzinska:2016ofb}.

\subsection{Numerical grid and determination of stellar surface}

The stellar models are constructed on a numerical grid where the
computational domain in spherical polar coordinates covers the regions
$0 \leq r \leq \infty$ and $0 \leq \theta \leq 2\pi$. Instead of the
coordinates $(r,\theta)$ in Equation ~\eqref{eq:ds}, the code solves the
coupled Einstein-hydrostationary equations in coordinates defined by
$u=\cos\theta$, and a compactified radial coordinate $s$ that maps
spatial infinity onto the computational domain as
\begin{equation}\label{eq:auxil_r}
r \equiv r_e \left(\dfrac{s}{1-s}\right).
\end{equation}
By construction, the surface of the star on the
equator corresponds to $r \longrightarrow r_e$ and $s
\longrightarrow \frac{1}{2}$.

Adopting the coordinates $(u,s)$ results in the radial grid points
being concentrated closer to the origin. This is not very convenient,
because it does not allow an accurate determination of the stellar
surface, which is necessary to compute $\hat \beta$ through the
surface derivative appearing in Equation ~\eqref{eq:defbeta}.
To resolve this problem we adopt very high radial resolution. We use
linear interpolation along $r$ of the pressure ($p$) to determine the
location where the pressure drops to $10^{10} \rm dyn/cm^2$, which is
more than 20 orders of magnitude below the maximum pressure in the
neutron star models. We call that location the surface of the star. We
have experimented with higher order interpolation, too, but found that
linear interpolation exhibits convergence to within 1\% in most cases,
and within 3\% at most in some cases, in finding the surface at the
adopted radial resolutions. This is not the case with higher order
interpolation because it is oscillatory. This procedure determines the
surface of the star as $r_{\rm surf}(u)$, which
we use to compute numerically the derivative needed for
$\hat{\beta}$ in Eq.~\eqref{eq:defbeta}, which we re-express as
\begin{equation}\label{eq:beta_hat_derivs}
\hat{\beta} = - \left(\dfrac{r_e}{r_p}\right)^2 \left( \dfrac{\frac{dz^2}{du^2}}{\frac{d\varpi^2}{du^2}} \right)_{r_e},
\end{equation}
where 
\begin{equation}\label{eq:surf_z}
z^2 = [r_{\rm surf}(u)]^2 u^2
\end{equation}
and
\begin{equation}\label{eq:surf_rho}
\varpi^2 = [r_{\rm surf}(u)]^2(1 - u^2).
\end{equation}
We use a 3-point one-sided stencil for finite differencing combined with 
high radial resolution 
on the solution grid to determine the numerical derivatives in Equation~\eqref{eq:beta_hat_derivs}. We determine the necessary grid resolution by
 calculating $\hat{\beta}$ for benchmark sequences including spheroidal, 
 quasi-toroidal, and near mass-shedding models at increasing resolution 
 until the results converge to within 1\% accuracy in most cases, but within 3\% at most in some cases.
  A typical 
 configuration is constructed with 500 grid points covering the
 equatorial radius for polytropes, 1250 points covering the
 equatorial radius for nuclear EOSs, and 500 grid points covering the
 angular direction in all cases. 
 All parameters in the quadruplet $(\epsilon_{\rm max},
 \frac{r_p}{r_e}, \hat{A}^{-1}, \hat{\beta})$ besides $\hat{\beta}$
 are specified as inputs to the Cook code.

%

\subsection{Solution Space of a $\Gamma =2$ Polytrope}
 \begin{figure}[htb]
   \includegraphics[width=8.5cm]{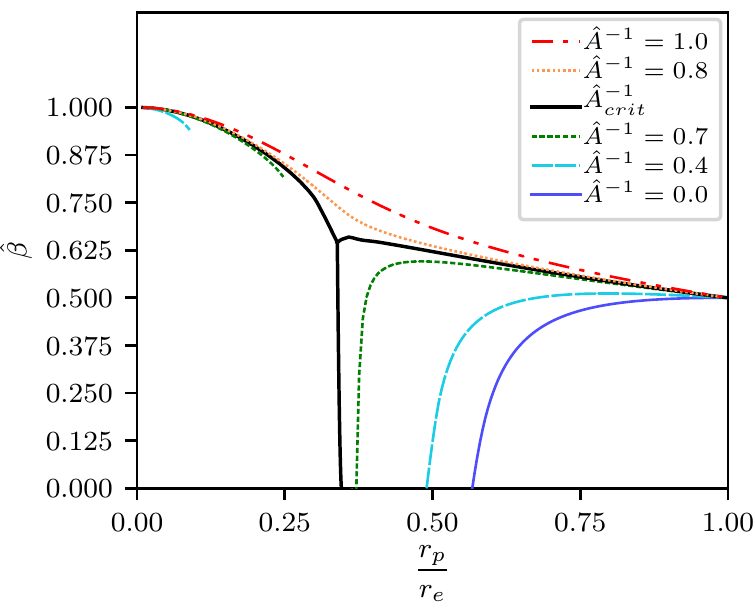}
   \caption{Mass-shedding parameter $\hat{\beta}$ as a
     function of $\frac{r_p}{r_e}$ at fixed maximum energy density
     $\epsilon_{\rm max}=0.12$ for a $\Gamma=2$ polytrope at varying
     degrees of differential rotation. The solid black line shows the
     separatrix at the critical value of differential rotation
     $\hat{A}^{-1}_{\rm crit} = 0.75904$ found in~\cite{Ansorg:2009},
which divides the solution space into the three regions wherein we are
     able to build equilibrium models. The colored lines show the
     characteristic sequences of equilibrium models for spheroids
     (Type A, right of the separatrix), quasi-toroids (Type B, left of
     the separatrix), and spheroids/quasi-toroids (Type C, above the
     separatrix). }
   \label{fig:solution_poly}
 \end{figure}
 
A polytropic EOS is described by 
\begin{equation}
p = \kappa \rho_0^\Gamma,
\end{equation}
where $p$ is the pressure, $\rho_0$ is the rest mass energy density, $\kappa$ is the polytropic
constant, and $\Gamma$ is the adiabatic index. When treating
polytropes, we employ polytropic units, such that $\kappa = G = c =
1$. For a $\Gamma=2$ polytrope,~\cite{Ansorg:2009}
and~\cite{Gondek-Rosinska:2016tzy} showed that there exist four
types of solutions, as we discussed in Sec.~\ref{sec:review}, and focused
on the maximum rest mass models obtainable for each type of
solution. In~\cite{Gondek-Rosinska:2016tzy} it was speculated
that~\cite{Lyford:2002ip} was unable to discover the different types
of solutions of differentially rotating stars due to limitations of
the Cook code. Here we demonstrate that the Cook code can reproduce
many of the $\Gamma=2$ results reported
in~\cite{Gondek-Rosinska:2016tzy}. We find that how one searches the
parameter space is the greatest limitation in constructing different
types of differentially rotating stars. Given that the code
of~\cite{Gondek-Rosinska:2016tzy} is spectral, we use the results
reported in that work to gauge the accuracy of the Cook code.

  \begin{table}[t]
 	\centering
 	\caption{Critical degree of differential rotation
          $\hat{A}^{-1}_{\rm crit}$ at several values of the maximum
          energy density $\epsilon_{\rm max}$ (and log of specific
          enthalpy $H_{\rm max}\equiv {\rm log}(h_{\rm max})$) in
          polytropic units for polytropes of four different polytropic
          indices $\Gamma$. Also shown is the percent error
          (calculated using \eqref{eq:res_err}) for each value of
          $\hat{A}^{-1}_{\rm crit}$ compared with those of Table A1 in
          \cite{Studzinska:2016ofb}.
 \label{tab:Acrits_gammas}}
\begin{tabular}{c c c | c c}\hline \hline 
$\Gamma$ & $\epsilon_{\rm max}$ & $H_{\rm max}$ & $\hat{A}^{-1}_{\rm crit}$ & $\delta\left(\hat{A}^{-1}_{\rm crit}\right)$\\ \hline
1.8 & 0.023 & 0.1 & 1.016 & 0.294\\
2.0 & 0.123 & 0.2 & 0.758 & 0.132\\ 
2.5 & 0.402 & 0.3 & 0.480 & 0.629\\
3.0 & 0.667 & 0.4 & 0.340 & 0.295\\ \hline
\end{tabular}
 \end{table} 
Unlike the code of~\cite{Gondek-Rosinska:2016tzy} which employs
surface fitted grids and also appears to be able to control the
parameter $\hat\beta$, the Cook code builds rotating stars by
specifying the triplet $(\epsilon_{\rm max}, \frac{r_p}{r_e},
\hat{A}^{-1})$. Once a configuration has been built, $\hat{\beta}$ is
determined by use of Eq.~\eqref{eq:beta_hat_derivs}. This makes
scanning the full parameter space challenging, and is probably the
reason why we were not able to build sequences of Type D and
lower-$\hat\beta$ Type B stars.

At a given value of $\epsilon_{\rm max}$, there exists a critical
degree of differential rotation at which the solution space exhibits
equilibrium solutions of all types (A, B, C, and D). Three out of the
four solution types we were able to construct with the Cook code for
$\epsilon_{\rm max}=0.12$ are shown in Figure
~\ref{fig:solution_poly}.  The solid black curve in the plot is the
separatrix in the solution space which corresponds to the critical
degree of differential rotation, and separates the space into four
regions, each corresponding to a solution type (although here we have
only three regions because we could not generate Type D
sequences). Type A solutions are found on the lower right part of the
plot, e.g., with values of $\hat{A}^{-1} \in \{0.0, 0.4, 0.7\}$; Type B
  solutions are found on the left side of the plot, e.g., with values of
  $\hat{A}^{-1} \mathbb{\in} \{0.4, 0.7\}$; the Type C solutions are
  found along the top part of the plot, e.g., with values of $\hat{A}^{-1} \in
  \{0.8, 1.0\}$.

 \begin{table*}[htb]
  \centering
  \caption{Listed are the degree of differential rotation $\hat{A}^{-1}$,
    ratio of polar to equatorial radius $\frac{r_p}{r_e}$, and maximum
    energy density $\epsilon_{\rm max}$ for the maximum rest mass
    models of a $\Gamma=2$ polytrope. Also shown for each model are
    the ratio of central to equatorial angular velocity
    $\frac{\Omega_c}{\Omega_e}$, the rest mass $M_0$, the ratio of
    kinetic to potential energy $\frac{T}{\vert W \rvert}$, angular
    momentum $J$, and ratio of ADM mass to circumferential radius
    $\frac{M}{R_c}$. For each quantity of interest we also report the percent error [$\delta()$] as defined in Equation ~(\ref{eq:res_err}). \label{tab:maxmass_poly}}
  \begin{tabular}{c | c c c | c c | c c | c c | c c | c c }\hline  \hline
Type & $\hat{A}^{-1}$ & $\dfrac{r_p}{r_e}$ & $\epsilon_{\rm max}$ & $\dfrac{\Omega_c}{\Omega_e}$ & $\delta\left(\dfrac{\Omega_c}{\Omega_e}\right)$ & $M_0$ & $\delta M_0$ & $ \dfrac{T}{|W|}$ & $\delta\left(\dfrac{T}{|W|}\right)$ & $J$ & $\delta J$ & $\dfrac{M}{R_c}$ & $\delta\left(\dfrac{M}{R_c}\right)$ \\ \hline
A & 0.0 & 0.585 &  0.350 & 1.000 & 0.000 & 0.207 & 0.029 & 0.083 & 0.240 & 0.020 & 0.843 & 0.174 & 0.155 \\
& 0.1 & 0.580 &  0.349 & 1.027 & 0.000 & 0.208 & 0.037 & 0.086 & 0.467 & 0.021 & 1.597 & 0.174 & 0.040 \\
& 0.2 & 0.565 &  0.347 & 1.108 & 0.000 & 0.211 & 0.037 & 0.093 & 0.432 & 0.022 & 0.677 & 0.174 & 0.275 \\
& 0.3 & 0.541 &  0.343 & 1.240 & 0.000 & 0.216 & 0.055 & 0.104 & 0.192 & 0.025 & 1.215 & 0.176 & 0.245 \\
& 0.4 & 0.511 &  0.335 & 1.422 & 0.000 & 0.224 & 0.011 & 0.120 & 0.332 & 0.028 & 1.720 & 0.178 & 0.231 \\
& 0.5 & 0.473 &  0.323 & 1.657 & 0.000 & 0.236 & 0.183 & 0.142 & 0.070 & 0.034 & 0.176 & 0.181 & 0.121 \\
& 0.6 & 0.427 &  0.304 & 1.959 & 0.000 & 0.254 & 0.079 & 0.171 & 0.117 & 0.043 & 0.327 & 0.188 & 0.181 \\
& 0.7 & 0.352 &  0.306 & 2.518 & 0.439 & 0.294 & 0.396 & 0.222 & 0.090 & 0.062 & 0.689 & 0.221 & 3.107 \\ \hline
B & 0.4 & 0.035 &  0.089 & 1.774 & 0.616 & 0.682 & 5.409 & 0.331 & 1.488 & 0.381 & 9.716 & 0.280 & 3.704 \\
& 0.5 & 0.114 &  0.084 & 1.976 & 1.496 & 0.586 & 8.294 & 0.324 & 3.284 & 0.289 & 15.000 & 0.259 & 5.285 \\
& 0.6 & 0.144 &  0.081 & 2.196 & 1.215 & 0.516 & 9.632 & 0.313 & 5.438 & 0.227 & 18.051 & 0.242 & 9.009 \\
& 0.7 & 0.164 &  0.081 & 2.458 & 0.614 & 0.463 & 9.216 & 0.302 & 6.790 & 0.184 & 18.222 & 0.231 & 14.925 \\ \hline
C & 0.8 & 0.005 &  0.097 & 2.997 & 0.067 & 0.463 & 0.041 & 0.294 & 0.102 & 0.176 & 0.114 & 0.250 & 0.160 \\
& 0.9 & 0.002 &  0.100 & 3.388 & 0.177 & 0.434 & 0.099 & 0.285 & 0.140 & 0.152 & 0.393 & 0.246 & 0.408 \\
& 1.0 & 0.005 &  0.103 & 3.809 & 0.105 & 0.409 & 0.120 & 0.277 & 0.036 & 0.134 & 0.149 & 0.241 & 0.207 \\
& 1.5 & 0.010 &  0.121 & 6.431 & 0.171 & 0.326 & 0.031 & 0.238 & 0.042 & 0.079 & 0.894 & 0.228 & 0.220 \\ \hline 
\end{tabular}
 \end{table*}
 
 \subsection{Solution space}
 \begin{figure*}[htb]
 	\centering
\begin{tabular}{c c}
\includegraphics[width=8.5cm]{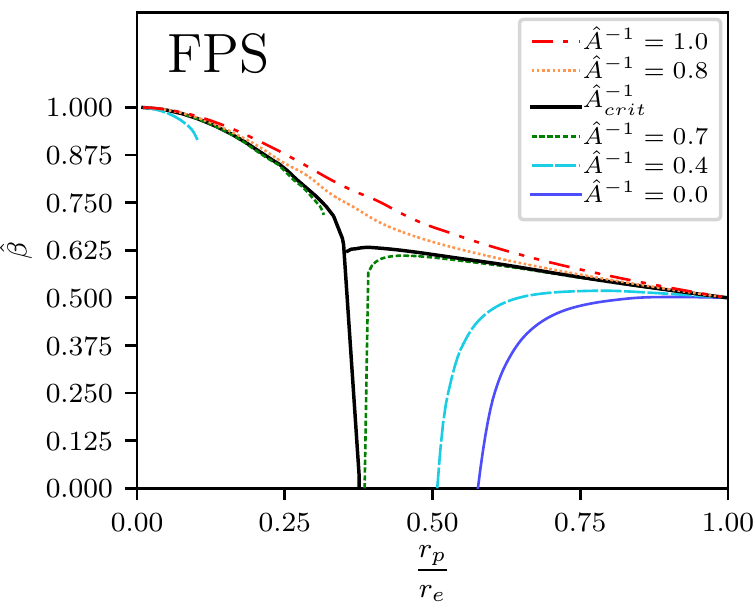}
\includegraphics[width=8.5cm]{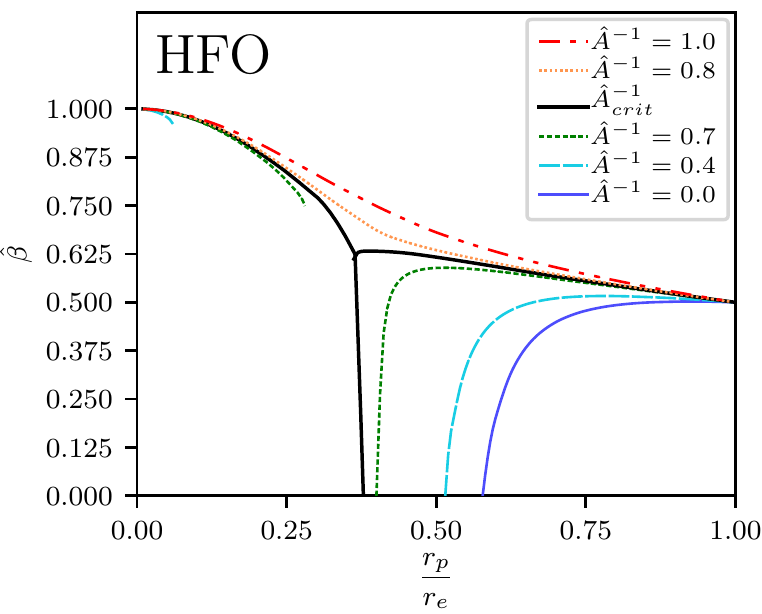}\\
\includegraphics[width=8.5cm]{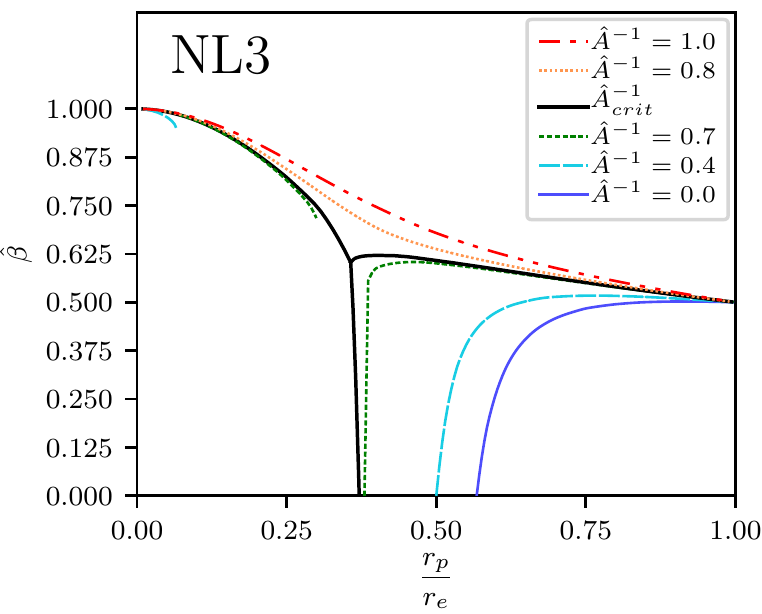}\includegraphics[width=8.5cm]{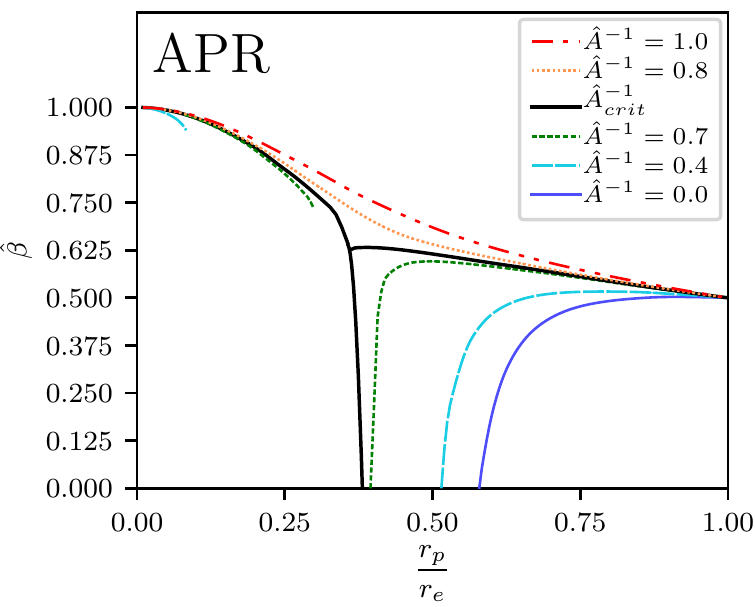}\\
\end{tabular}
 \caption{Solution space for the FPS, HFO, NL3 and APR EOSs.  These
   plots correspond to fixed energy densities, the values of which,
   along with $\hat{A}^{-1}_{\rm crit}$, are displayed in
   Table~\ref{tab:solution_densities_realistic} for each
   EOS.}\label{fig:solutions_realistic}
 \end{figure*}  
 
It was shown in~\cite{Ansorg:2009} that for a fixed value
of $\epsilon_{\rm max}$, $\hat{A}^{-1}$ as a function of
$\frac{r_p}{r_e}$ and $\hat{\beta}$ exhibits a saddle point at the
value $\hat{A}^{-1} = \hat{A}^{-1}_{\rm crit}$, so that the solution to
the equations
\begin{equation}\label{eq:Acrit_extremum}
  \left(\dfrac{\partial \hat{A}^{-1}}{\partial(r_p/r_e)} \right)_{\epsilon_{\rm max}}= 0 = \left(\dfrac{\partial \hat{A}^{-1}}{\partial\hat{\beta}} \right)_{\epsilon_{\rm max}}
\end{equation}
defines the value $\hat{A}^{-1}_{\rm crit}$. Instead of solving these
equations we use a different method to find the critical degree of
differential rotation.  For each $\frac{r_p}{r_e}$ and at fixed
$\epsilon_{\rm max}$, there exists a minimum value of $\hat{A}^{-1}$
for which equilibrium solutions exist. We denote this minimum value
$\hat{A}^{-1}_{\rm min}$. The function $\hat{A}^{-1}_{\rm
  min}(\frac{r_p}{r_e})$ exhibits a maximum, and the maximum value is
$\hat{A}^{-1}_{\rm crit}$.  We effectively solve Equation
\eqref{eq:Acrit_extremum} by locating the maximum in the
$\left(\frac{r_p}{r_e}, \hat{A}^{-1}_{\rm min}\right)$ plane. We find
that this extremum is a \textit{global} maximum, so that it is
possible to accurately locate the value of $\hat{A}^{-1}_{\rm crit}$
with our method instead of actually solving
Eq.~\eqref{eq:Acrit_extremum} as was done
in~\cite{Gondek-Rosinska:2016tzy}.  The critical value we find for
$\Gamma=2$ polytropes at $\epsilon_{\rm max}=0.12$ is
$\hat{A}^{-1}_{\rm crit}=0.7612$, which is only $\sim 0.284 \% $
greater than the critical value of $\hat{A}^{-1}_{\rm crit}=0.75904$
found in~\cite{Gondek-Rosinska:2016tzy}.  To more accurately determine
the value of $\hat{A}_{\rm crit}^{-1}$ we slowly lower the value of
$\hat{A}^{-1}$ until we reach a value that exhibits solutions of all
types (except for type D which we cannot build), which are
continuously joined (the defining feature of the separatrix). This
procedure allows for the determination of $\hat{A}_{\rm crit}^{-1}$ to
better than $1\%$ accuracy. In Table \ref{tab:Acrits_gammas} we show
the value of $\hat{A}^{-1}_{\rm crit}$ found using our method for
polytropes accross several polytropic indices and values of the
maximum energy density. Values of $\hat{A}^{-1}_{\rm crit}$ at the
same maximum energy densities and for the same polytropic indices can
be found in Table A1 in \cite{Studzinska:2016ofb}. All of the values
of $\hat{A}^{-1}_{\rm crit}$ presented in Tab. \ref{tab:Acrits_gammas}
agree with those in Table A1 in \cite{Studzinska:2016ofb} to within
1\%. Note that we also list the logarithm of the specific enthalpy
$H_{\rm max} \equiv {\rm log}(h_{\rm max})$, where
\begin{equation}\label{eq:spec_enth}
h = \dfrac{\epsilon + p}{\rho_0},
\end{equation}
to offer easier comparison to the results of \cite{Studzinska:2016ofb}. 

In Table~\ref{tab:maxmass_poly} we show properties of the maximum rest mass models
obtained for a $\Gamma=2$ polytrope. For
each quantity also computed in \cite{Gondek-Rosinska:2016tzy}, we show
the percent error between our models and the corresponding
ones in~\cite{Gondek-Rosinska:2016tzy}, computed as
\begin{equation}\label{eq:res_err}
\delta x \equiv \dfrac{\lvert x - x_{\rm ref} \rvert}{x_{\rm ref}}\times 100,
\end{equation}
where $x$ represents the values obtained using the Cook code, and
$x_{\rm ref}$ represents the values presented in
\cite{Gondek-Rosinska:2016tzy}. Given that the code
of~\cite{Gondek-Rosinska:2016tzy} is spectral, $\delta x$ is an
estimate of the error in our calculations for the
resolution we adopt. Note that we also show $\delta x$ for the values of $\hat{A}^{-1}_{\rm crit}$ in Table \ref{tab:Acrits_gammas}.

The highest fractional differences are seen in the Type B models,
going as high as $\mathcal{O}(10\%)$ in the rest mass and angular
momentum for the most massive configuration and less than $10\%$ in
other quantities; in all other cases the errors are sub-percent. We
suspect that the relatively high residuals in some of the Type B
models we built are due to the fact that the solutions presented in
\cite{Gondek-Rosinska:2016tzy} are near the mass-shedding limit
(highly pinched and quasi-toroidal at low $\hat{\beta}$), whereas the
corresponding models presented here belong to the part of the Type B
sequence at higher values of $\hat{\beta}$. Close inspection of the Type
B sequences in \cite{Gondek-Rosinska:2016tzy} shows that they are not
always single-valued in $\frac{r_p}{r_e}$, so that without the full
solution space coordinates (i.e, the full quadruplet $(\epsilon_{\rm
  max}, \frac{r_p}{r_e}, \hat{A}^{-1},\hat{\beta})$) two distinct
models may be misindentified as the same equilibrium solution. Because
only the triplet $(\epsilon_{\rm max}, \frac{r_p}{r_e}, \hat{A}^{-1})$
is presented in \cite{Gondek-Rosinska:2016tzy} for these maxmimum mass
models, we cannot be sure that we are comparing the same two
models. However, the confidence in our solutions is supported by the
fact that the majority of other cases show sub-percent residuals in
all of the model properties.

The highest mass models built in \cite{Gondek-Rosinska:2016tzy} and
\cite{Studzinska:2016ofb} were of Type B with the lowest value of
$\hat{A}^{-1}$ among those considered. We note that the maximum rest
mass Type D models presented in \cite{Gondek-Rosinska:2016tzy} and
\cite{Studzinska:2016ofb} neither exceed the maximum rest mass Type B
models nor Type C models in the rest mass in all cases where
they could be built. We anticipate that this result holds true for
realistic EOSs, too.  Although we were unable to construct a suitable
sequence of Type D models with the Cook code, Type D models are
likely unphysical as pointed out
in~\cite{Studzinska:2016ofb}. Despite the limitations of non-spectral
codes, here we showed that the Cook code can generate Type A, B, and C
models, and closely match the maximum-mass configurations for a
$\Gamma=2$ polytrope obtained with a spectral code. This result gives us confidence that the maximum-mass models we report for
realistic EOSs in the next section are the true maximum-mass type A and C modes, and very close to the true maximum-mass Type B models.

\section{Results with realistic equations of state}
\label{sec:results}
   \begin{figure*}[htp]
 	\centering
\begin{tabular}{c c}
\includegraphics[width=8.5cm]{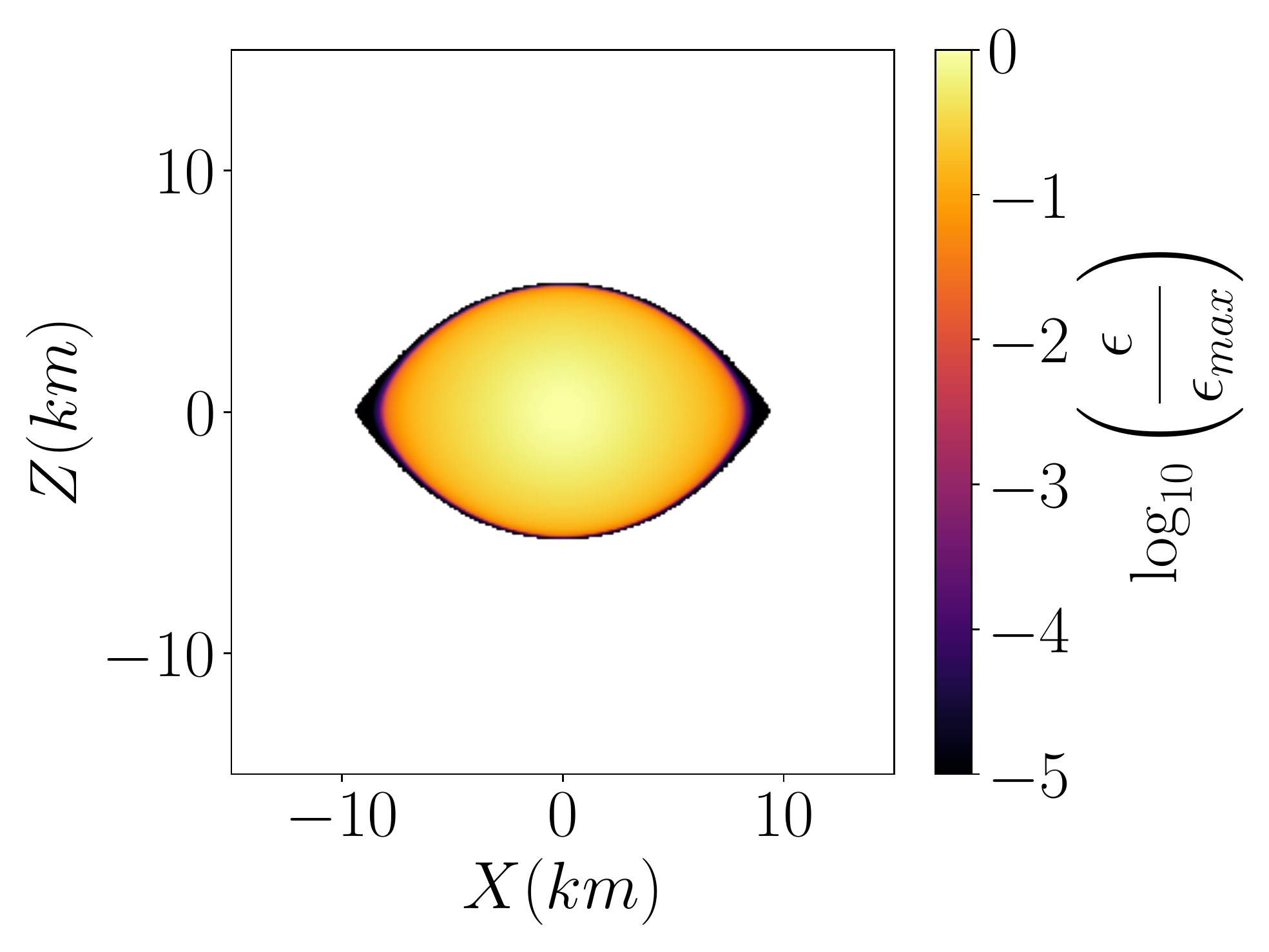}\includegraphics[width=8.5cm]{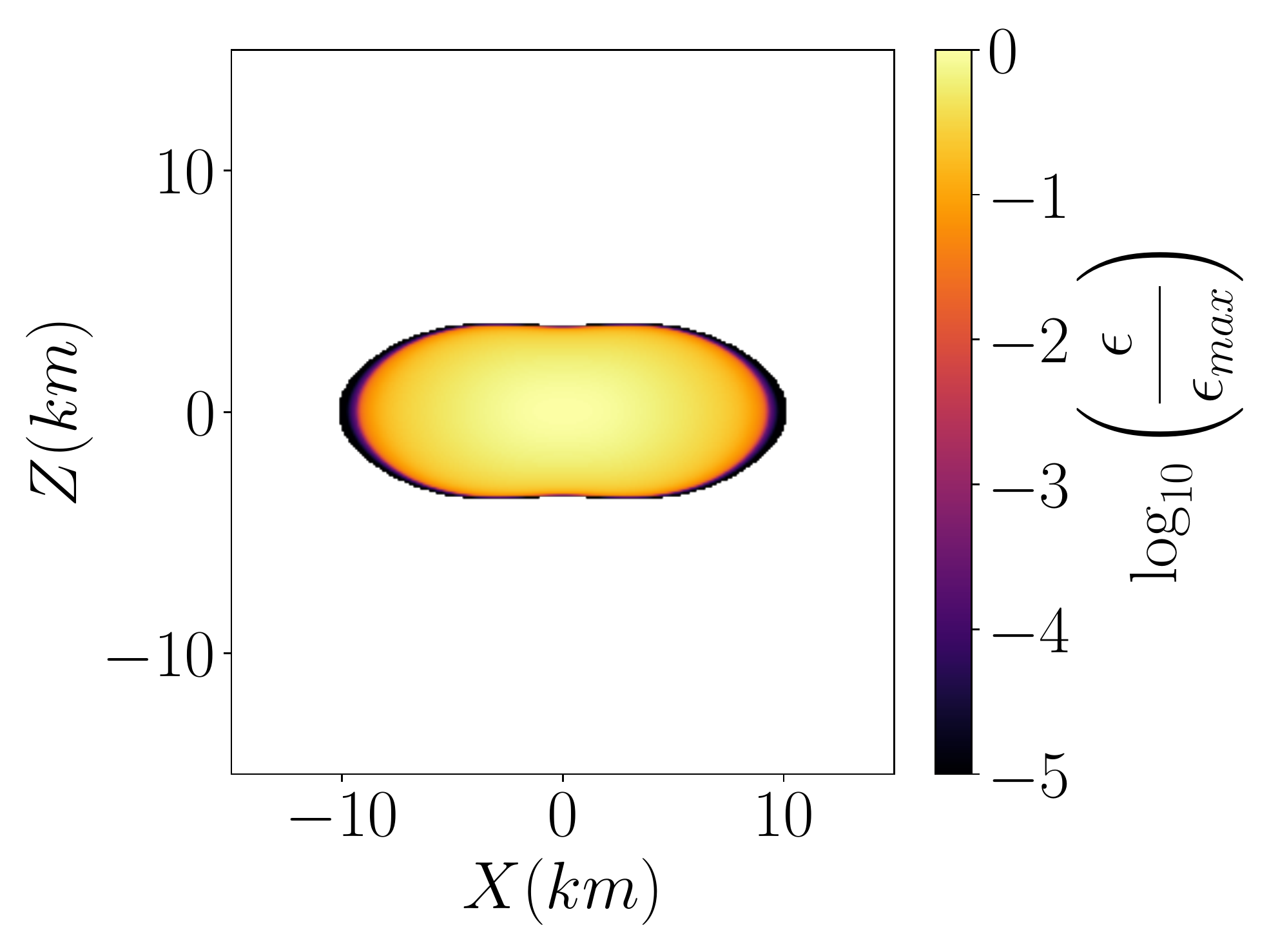}\\
\includegraphics[width=8.5cm]{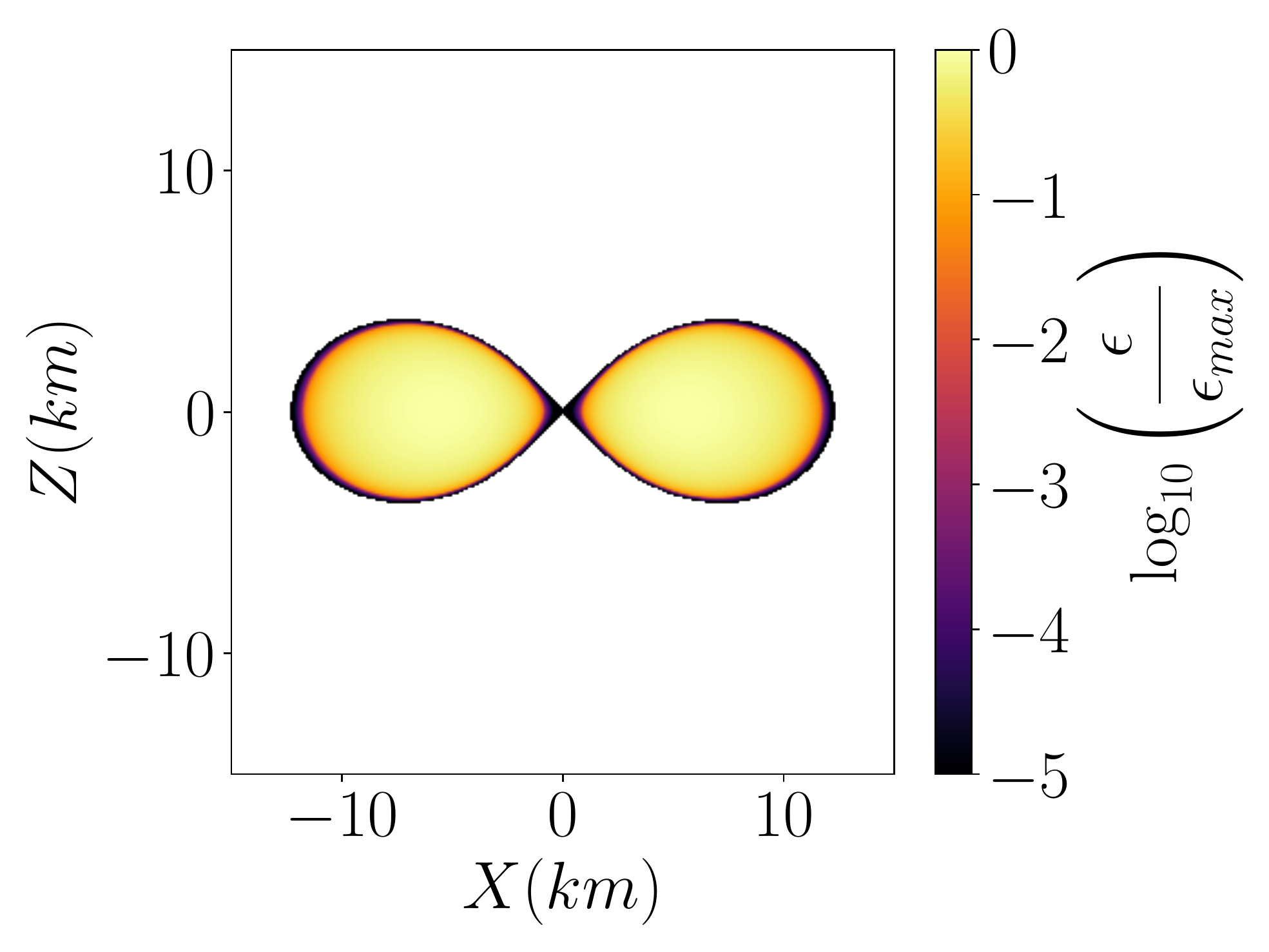}\includegraphics[width=8.5cm]{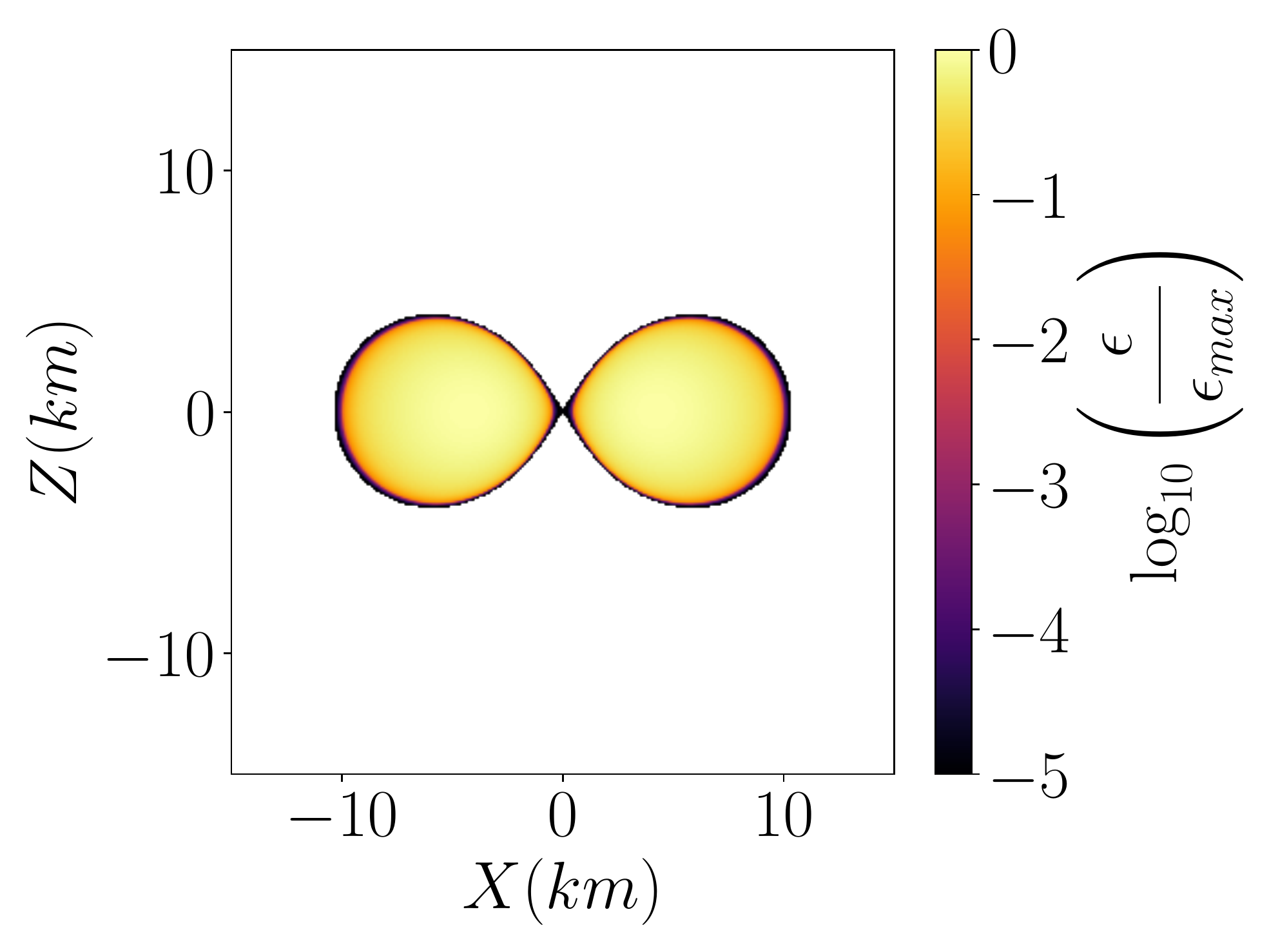}\\
\end{tabular}
 \caption{Examples of meridional energy density contours for the HFO
   EOS. Top left: the maximum rest mass uniformly rotating
   ($\hat{A}^{-1} = 0.0$). Top right: The maximum rest mass Type A
   model. Bottom left: the maximum rest mass Type B model. Bottom
   right: the maximum rest mass Type C model.
 } \label{fig:meridional} 
 \end{figure*}
 
  \begin{table}[t]
 	\centering
 	\caption{Maximum energy density $\epsilon_{\rm max}$ and corresponding critical degree of differential rotation $\hat{A}^{-1}_{\rm crit}$ used in generating the solution spaces shown in Fig.~\ref{fig:solutions_realistic} for realistic EOSs. \label{tab:solution_densities_realistic}}
\begin{tabular}{c c c}\hline \hline 
EOS & $\dfrac{\epsilon_{\rm max}}{10^{15}\rm g/cm^3}$ & $\hat{A}^{-1}_{\rm crit}$\\ \hline
FPS & 0.77 & 0.7161\\
HFO & 0.6 & 0.753\\ 
NL3 & 0.35 & 0.717\\
APR & 0.7 & 0.7376\\ \hline
\end{tabular}
 \end{table} 
 
In this section we discuss the solution space of differentially
rotating, relativistic stars with realistic equations of state and the
maximum rest mass they can support.

The solution space depends on the value of $\epsilon_{\rm max}$. To
reveal as large a fraction of the space of solutions as possible, for
each EOS we obtain the critical degree of differential rotation
$\hat{A}^{-1}_{\rm crit}$ for different values of $\epsilon_{\rm
  max}$. Then we choose the values of $\epsilon_{\rm max}$ for which
$0.7 \leq \hat{A}^{-1}_{\rm crit} \leq 0.8$. With this choice of
$\epsilon_{\rm max}$, three out of four types of sequences we are able
to construct are present for each of the EOSs considered. Moreover,
models with $\hat{A}^{-1}\in [0.0,0.4,0.7]$ belong to
sequences of Type A and B, and models with $\hat{A}^{-1}\in
[0.8,1.0]$ 
belong to sequences of Type C.
As in the $\Gamma=2$ polytrope in the
previous section, to scan the parameter space we fix the value of
$\epsilon_{\rm max}$, modify the parameters $(\hat{A}^{-1},
\frac{r_p}{r_e})$ to construct stellar models and compute
$\hat{\beta}$. Our results for the solution space of realistic EOSs
with differential rotation at fixed $\epsilon_{\rm max}$ are shown in Figure~\ref{fig:solutions_realistic}. The values of $\epsilon_{\rm max}$
and $\hat{A}^{-1}_{\rm crit}$ for each EOS that correspond to
Fig.~\ref{fig:solutions_realistic} are given in Table
\ref{tab:solution_densities_realistic}. Fig.~\ref{fig:solutions_realistic} demonstrates that the existence
of different types of differentially rotating stars are not a property
of polytropic EOSs only. The different types exist for realistic EOSs,
too.

In Figure~\ref{fig:meridional}, we show meridional contours of the
energy density $\epsilon$ normalized to $\epsilon_{\rm max}$ for
different types of differentially rotating stars constructed with the
HFO EOS.  The top left and right panels of Fig.~\ref{fig:meridional}
depict the maximum rest mass Type A models for $\hat{A}^{-1} = 0.0$
(uniform rotation) and $\hat{A}^{-1} = 0.4$ (largest rest mass Type A
model), respectively. The bottom left and bottom right panels depict
the largest rest mass Type B and Type C models, respectively.

Although we were not able to build complete sequences of Type D
equilibria, we were able to construct individual \textit{candidate}
configurations near the mass-shedding limit, and for values of
$\hat{A}^{-1} > \hat{A}^{-1}_{\rm crit}$, all of which are properties of
Type D models. For example, one candidate Type D configuration for the HFO EOS corresponds to
$\epsilon_{\rm max} = 6\times 10^{14} \rm g/cm^3$, $\frac{r_p}{r_e} =
0.375$, $\hat{A}^{-1} = 0.757$, and $\hat{\beta} = 0.064$. This candidate Type D model has $\dfrac{M_0}{M_{0, \rm max}^{TOV}} = 0.574$ and $\dfrac{M_0}{M_{0, \rm max}^{sup}} = 0.464$, meaning that they are less massive than the maximum rest mass TOV model of HFO. We were
able to construct this model by finding a model close to mass-shedding
along the separatrix for the panel corresponding to HFO in
Fig.~\ref{fig:solutions_realistic} (i.e, using the values of
$\epsilon_{\rm max}$ and $\hat{A}^{-1}$ from Table
\ref{tab:maxmass_HO} for HFO). Once the closest model to mass-shedding
for the separatrix was built, we decreased the value of
$\frac{r_p}{r_e}$ while increasing the value of $\hat{A}^{-1}$ and
searched for models near mass-shedding.

\subsection{Maximum rest mass}
\label{sec:discussion}
\begin{table*}[htp] 
  \centering 
    \caption{
        Maximum rest mass models for the FPS EOS. Shown are the values of the degree
      of differential rotation $\hat{A}^{-1}$, the maximum energy
      density $\epsilon_{\rm max}$ in units of $10^{15}\dfrac{\rm g}{\rm cm^3}$, the ratio of polar to equatorial
      radius $\frac{r_p}{r_e}$, the mass-shedding parameter
      $\hat{\beta}$, the circumferential radius $R_c$ in units of km, the ratio of
      kinetic to gravitational potential energy $\frac{T}{\lvert W
        \rvert}$, the ratio of central to equatorial angular velocity
      $\frac{\Omega_c}{\Omega_e}$, the dimensionless spin
      $\frac{J}{M^2}$, the compactness $C = \frac{M_{\rm ADM}}{R_c}$, the rest mass $M_0$, the ratio of rest mass to
      the TOV limit rest mass $M_{0, \rm max}^{TOV}$, the ratio of rest
      mass to the supramassive limit rest mass $M_{0, \rm max}^{sup}$, and
      the ADM mass $M_{\rm ADM}$ along with the ratio of ADM mass to
      TOV limit ADM mass $M_{\rm ADM,max}^{TOV}$ and the supramassive
      limit ADM mass $M_{\rm ADM,max}^{sup}$. Also shown is the classification of each star as supramassive (SUP), hypermassive (HYP) or \"ubermassive (\"UBE). }
  \label{tab:maxmass_FPS} 
{\begin{tabular}{  c | c | c | c | c | c | c | c | c | c | c c c | c c c | c }\hline \hline
Type & $\hat{A}^{-1}$ & $\epsilon_{\rm max}$ & $\frac{r_p}{r_e}$ & $\hat{\beta}$ & $R_c$ & $\frac{T}{|W|}$ & $\frac{\Omega_c}{\Omega_e}$ & $\frac{J}{M^2}$ & $C$ & $\frac{M_0}{M_\odot}$ & $\frac{M_0}{M_{0, \rm max}^{TOV}} $ & $\frac{M_0}{M_{0, \rm max}^{sup}} $ & $\frac{M_{\rm ADM}}{M_\odot}$ & $\frac{M_{\rm ADM}}{M_{\rm ADM, max}^{TOV}}$ & $\frac{M_{ADM}}{M_{\rm ADM,max}^{sup}}$ & CLASS \\ \hline
A & 0.0 & 2.92 &  0.568 & 0.060 &  12.44 & 0.117 & 1.000 & 0.658 & 0.170 & 2.452 & 1.167 & 1.000 & 2.120 & 1.178 & 1.000 & SUP\\
& 0.1 & 2.92 &  0.557 & 0.097 &  12.50 & 0.123 & 1.049 & 0.674 & 0.171 & 2.478 & 1.179 & 1.010 & 2.143 & 1.190 & 1.011 & HYP\\
& 0.2 & 2.90 &  0.526 & 0.202 &  12.69 & 0.141 & 1.194 & 0.719 & 0.174 & 2.557 & 1.217 & 1.043 & 2.213 & 1.230 & 1.044 & HYP\\
& 0.3 & 2.87 &  0.470 & 0.305 &  13.05 & 0.173 & 1.442 & 0.786 & 0.180 & 2.713 & 1.291 & 1.106 & 2.350 & 1.305 & 1.108 & HYP\\
& 0.4 & 2.47 &  0.387 & 0.480 &  13.73 & 0.226 & 1.839 & 0.874 & 0.193 & 3.059 & 1.455 & 1.247 & 2.648 & 1.471 & 1.249 & HYP\\
& 0.5 & 1.34 &  0.361 & 0.547 &  16.43 & 0.241 & 1.888 & 0.923 & 0.159 & 2.983 & 1.419 & 1.217 & 2.619 & 1.455 & 1.236 & HYP\\ \hline
B & 0.4 & 0.94 &  0.010 & 1.000 &  20.65 & 0.327 & 1.988 & 1.026 & 0.217 & 5.238 & 2.492 & 2.136 & 4.490 & 2.494 & 2.118 & \"UBE\\
& 0.5 & 0.98 &  0.010 & 1.000 &  19.10 & 0.317 & 2.335 & 1.009 & 0.212 & 4.699 & 2.235 & 1.916 & 4.054 & 2.252 & 1.912 &\"UBE
\\ \hline
C & 0.6 & 1.01 &  0.010 & 1.000 &  18.010 & 0.306 & 2.705 & 0.991 & 0.207 & 4.297 & 2.044 & 1.752 & 3.728 & 2.071 & 1.758 &\"UBE\\
& 0.7 & 1.05 &  0.010 & 1.000 &  17.040 & 0.295 & 3.132 & 0.970 & 0.203 & 3.983 & 1.895 & 1.624 & 3.467 & 1.926 & 1.636 &HYP\\
& 0.8 & 1.09 &  0.010 & 1.000 &  16.230 & 0.284 & 3.598 & 0.949 & 0.201 & 3.732 & 1.775 & 1.522 & 3.257 & 1.809 & 1.536 &HYP\\
& 0.9 & 1.14 &  0.010 & 1.000 &  15.490 & 0.273 & 4.127 & 0.926 & 0.199 & 3.526 & 1.678 & 1.438 & 3.083 & 1.712 & 1.454 &HYP\\
& 1.0 & 1.19 &  0.010 & 1.000 &  14.870 & 0.262 & 4.699 & 0.902 & 0.198 & 3.356 & 1.597 & 1.369 & 2.937 & 1.632 & 1.386 &HYP\\
& 1.5 & 1.47 &  0.010 & 0.990 &  12.720 & 0.210 & 8.329 & 0.788 & 0.195 & 2.828 & 1.345 & 1.153 & 2.478 & 1.377 & 1.169 &HYP\\ \hline
\end{tabular}}
\end{table*}
\begin{table*}[htp] 
  \centering 
    \caption{The columns list the same quantities as in Tab.~\ref{tab:maxmass_FPS} but for the HFO EOS.}
  \label{tab:maxmass_HO} 
{\begin{tabular}{  c | c | c | c | c | c | c | c | c | c | c c c | c c c | c }\hline \hline
Type & $\hat{A}^{-1}$ & $\epsilon_{\rm max}$ & $\frac{r_p}{r_e}$ & $\hat{\beta}$ & $R_c$ & $\frac{T}{|W|}$ & $\frac{\Omega_c}{\Omega_e}$ & $\frac{J}{M^2}$ & $C$ & $\frac{M_0}{M_\odot}$ & $\frac{M_0}{M_{0, \rm max}^{TOV}} $ & $\frac{M_0}{M_{0, \rm max}^{sup}} $ & $\frac{M_{\rm ADM}}{M_\odot}$ & $\frac{M_{\rm ADM}}{M_{\rm ADM, max}^{TOV}}$ & $\frac{M_{ADM}}{M_{\rm ADM,max}^{sup}}$ & CLASS \\ \hline
A & 0.0 & 2.32 &  0.564 & 0.078 &  13.710 & 0.125 & 1.000 & 0.677 & 0.178 & 2.829 & 1.174 & 1.000 & 2.440 & 1.187 & 1.000 & SUP\\
& 0.1 & 2.32 &  0.550 & 0.089 &  13.810 & 0.132 & 1.054 & 0.695 & 0.179 & 2.863 & 1.188 & 1.012 & 2.470 & 1.202 & 1.012 & HYP\\
& 0.2 & 2.30 &  0.515 & 0.207 &  14.040 & 0.153 & 1.214 & 0.746 & 0.183 & 2.972 & 1.234 & 1.051 & 2.567 & 1.249 & 1.052 & HYP\\
& 0.3 & 2.24 &  0.450 & 0.334 &  14.470 & 0.192 & 1.496 & 0.820 & 0.191 & 3.199 & 1.328 & 1.131 & 2.767 & 1.346 & 1.134 & HYP\\
& 0.4 & 1.54 &  0.376 & 0.565 &  15.690 & 0.245 & 1.869 & 0.900 & 0.200 & 3.624 & 1.504 & 1.281 & 3.134 & 1.525 & 1.284 & HYP\\
& 0.5 & 0.99 &  0.360 & 0.544 &  18.320 & 0.242 & 1.841 & 0.935 & 0.154 & 3.177 & 1.319 & 1.123 & 2.817 & 1.370 & 1.154 & HYP\\ \hline
B & 0.4 & 0.80 &  0.011 & 0.999 &  21.960 & 0.327 & 2.025 & 1.020 & 0.221 & 5.642 & 2.342 & 1.994 & 4.854 & 2.362 & 1.989 & \"UBE\\
& 0.5 & 0.83 &  0.010 & 1.000 &  20.380 & 0.316 & 2.378 & 1.002 & 0.215 & 5.070 & 2.105 & 1.792 & 4.391 & 2.136 & 1.800 &\"UBE\\ 
 \hline
C & 0.6 & 0.86 &  0.010 & 1.000 &  19.150 & 0.305 & 2.772 & 0.983 & 0.211 & 4.644 & 1.928 & 1.642 & 4.041 & 1.966 & 1.656 & HYP\\
& 0.7 & 0.90 &  0.010 & 1.000 &  18.060 & 0.293 & 3.231 & 0.960 & 0.208 & 4.312 & 1.790 & 1.524 & 3.763 & 1.831 & 1.542 & HYP\\
& 0.8 & 0.94 &  0.010 & 1.000 &  17.160 & 0.282 & 3.734 & 0.937 & 0.206 & 4.048 & 1.680 & 1.431 & 3.539 & 1.722 & 1.450 & HYP\\
& 0.9 & 0.99 &  0.010 & 1.000 &  16.330 & 0.270 & 4.314 & 0.911 & 0.205 & 3.833 & 1.591 & 1.355 & 3.355 & 1.632 & 1.375 & HYP\\
& 1.0 & 1.03 &  0.010 & 1.000 &  15.710 & 0.258 & 4.913 & 0.887 & 0.204 & 3.657 & 1.518 & 1.293 & 3.204 & 1.559 & 1.313 & HYP\\
& 1.5 & 1.29 &  0.010 & 1.000 &  13.440 & 0.203 & 8.863 & 0.769 & 0.203 & 3.120 & 1.295 & 1.103 & 2.732 & 1.329 & 1.120 & HYP\\ \hline
\end{tabular} }
\end{table*}
\begin{table*}[htp] 
  \centering 
  \caption{The columns list the same quantities as in Tab.~\ref{tab:maxmass_FPS} but for the NL3 EOS.}
  \label{tab:maxmass_G3} 
{\begin{tabular}{  c | c | c | c | c | c | c | c | c | c | c c c | c c c | c}\hline \hline
Type & $\hat{A}^{-1}$ & $\epsilon_{\rm max}$ & $\frac{r_p}{r_e}$ & $\hat{\beta}$ & $R_c$ & $\frac{T}{|W|}$ & $\frac{\Omega_c}{\Omega_e}$ & $\frac{J}{M^2}$ & $C$ & $\frac{M_0}{M_\odot}$ & $\frac{M_0}{M_{0, \rm max}^{TOV}} $ & $\frac{M_0}{M_{0, \rm max}^{sup}} $ & $\frac{M_{\rm ADM}}{M_\odot}$ & $\frac{M_{\rm ADM}}{M_{\rm ADM, max}^{TOV}}$ & $\frac{M_{ADM}}{M_{\rm ADM,max}^{sup}}$ & CLASS \\ \hline
A & 0.0 & 1.36 &  0.559 & 0.064 &  17.490 & 0.136 & 1.000 & 0.704 & 0.189 & 3.881 & 1.185 & 1.000 & 3.301 & 1.202 & 1.000 & SUP\\
& 0.1 & 1.36 &  0.540 & 0.080 &  17.690 & 0.145 & 1.062 & 0.726 & 0.190 & 3.940 & 1.203 & 1.015 & 3.353 & 1.221 & 1.016 & HYP\\
& 0.2 & 1.34 &  0.498 & 0.226 &  18.020 & 0.172 & 1.248 & 0.784 & 0.196 & 4.134 & 1.263 & 1.065 & 3.524 & 1.283 & 1.067 & HYP\\
& 0.3 & 1.24 &  0.411 & 0.395 &  18.780 & 0.225 & 1.603 & 0.872 & 0.209 & 4.598 & 1.405 & 1.185 & 3.925 & 1.429 & 1.189 & HYP\\
& 0.4 & 0.72 &  0.365 & 0.502 &  21.920 & 0.248 & 1.736 & 0.921 & 0.182 & 4.609 & 1.408 & 1.188 & 3.984 & 1.450 & 1.207 & HYP\\
& 0.5 & 0.54 &  0.359 & 0.521 &  23.820 & 0.242 & 1.754 & 0.957 & 0.141 & 3.777 & 1.154 & 0.973 & 3.367 & 1.226 & 1.020 & HYP\\ \hline
B & 0.4 & 0.50 &  0.010 & 1.000 &  27.180 & 0.326 & 2.050 & 1.016 & 0.224 & 7.114 & 2.173 & 1.833 & 6.082 & 2.214 & 1.842 & \"UBE\\
& 0.5 & 0.52 &  0.010 & 1.000 &  25.180 & 0.315 & 2.419 & 0.996 & 0.219 & 6.403 & 1.956 & 1.650 & 5.508 & 2.005 & 1.668 & HYP\\ \hline
C & 0.6 & 0.54 &  0.010 & 1.000 &  23.630 & 0.303 & 2.833 & 0.975 & 0.215 & 5.875 & 1.795 & 1.514 & 5.075 & 1.847 & 1.537 & HYP\\
& 0.7 & 0.57 &  0.010 & 1.000 &  22.140 & 0.290 & 3.342 & 0.948 & 0.214 & 5.468 & 1.670 & 1.409 & 4.731 & 1.722 & 1.433 & HYP\\
& 0.8 & 0.60 &  0.010 & 1.000 &  20.960 & 0.278 & 3.901 & 0.922 & 0.213 & 5.147 & 1.572 & 1.326 & 4.458 & 1.623 & 1.350 & HYP\\
& 0.9 & 0.63 &  0.010 & 1.000 &  19.980 & 0.265 & 4.515 & 0.895 & 0.212 & 4.889 & 1.493 & 1.260 & 4.237 & 1.542 & 1.283 & HYP\\
& 1.0 & 0.66 &  0.010 & 1.000 &  19.170 & 0.252 & 5.184 & 0.869 & 0.212 & 4.679 & 1.429 & 1.206 & 4.056 & 1.477 & 1.229 & HYP\\
& 1.5 & 0.83 &  0.010 & 1.000 &  16.500 & 0.193 & 9.511 & 0.746 & 0.213 & 4.059 & 1.240 & 1.046 & 3.509 & 1.277 & 1.063 & HYP\\ \hline
\end{tabular} }
\end{table*}
\begin{table*}[htp] 
  \centering
  \caption{The columns list the same quantities as in Tab.~\ref{tab:maxmass_FPS} but for the APR EOS.}
  \label{tab:maxmass_APR} 
{\begin{tabular}{  c | c | c | c | c | c | c | c | c | c | c c c | c c c | c}\hline \hline
Type & $\hat{A}^{-1}$ & $\epsilon_{\rm max}$ & $\frac{r_p}{r_e}$ & $\hat{\beta}$ & $R_c$ & $\frac{T}{|W|}$ & $\frac{\Omega_c}{\Omega_e}$ & $\frac{J}{M^2}$ & $C$ & $\frac{M_0}{M_\odot}$ & $\frac{M_0}{M_{0, \rm max}^{TOV}} $ & $\frac{M_0}{M_{0, \rm max}^{sup}} $ & $\frac{M_{\rm ADM}}{M_\odot}$ & $\frac{M_{\rm ADM}}{M_{\rm ADM, max}^{TOV}}$ & $\frac{M_{ADM}}{M_{\rm ADM,max}^{sup}}$ & CLASS \\ \hline
A & 0.0 & 2.42 &  0.564 & 0.059 &  12.900 & 0.137 & 1.000 & 0.709 & 0.201 & 3.091 & 1.163 & 1.000 & 2.599 & 1.187 & 1.000 & SUP\\
& 0.1 & 2.43 &  0.546 & 0.121 &  12.980 & 0.148 & 1.074 & 0.735 & 0.204 & 3.141 & 1.182 & 1.016 & 2.644 & 1.208 & 1.017 & HYP\\
& 0.2 & 2.41 &  0.490 & 0.248 &  13.270 & 0.181 & 1.298 & 0.801 & 0.210 & 3.306 & 1.244 & 1.070 & 2.793 & 1.276 & 1.075 & HYP\\
& 0.3 & 2.00 &  0.414 & 0.508 &  13.700 & 0.236 & 1.703 & 0.880 & 0.226 & 3.649 & 1.373 & 1.181 & 3.095 & 1.414 & 1.191 & HYP\\
& 0.4 & 1.27 &  0.368 & 0.510 &  16.360 & 0.247 & 1.769 & 0.912 & 0.186 & 3.547 & 1.335 & 1.148 & 3.047 & 1.392 & 1.172 & HYP\\
& 0.5 & 0.99 &  0.377 & 0.521 &  17.520 & 0.231 & 1.772 & 0.925 & 0.144 & 2.852 & 1.073 & 0.923 & 2.525 & 1.154 & 0.972 & HYP\\ \hline
B & 0.4 & 0.86 &  0.011 & 1.000 &  20.900 & 0.327 & 2.025 & 1.020 & 0.221 & 5.410 & 2.036 & 1.751 & 4.621 & 2.111 & 1.778 & \"UBE\\
& 0.5 & 0.91 &  0.010 & 1.000 &  19.090 & 0.315 & 2.432 & 0.996 & 0.219 & 4.875 & 1.835 & 1.578 & 4.182 & 1.910 & 1.609 & HYP\\ \hline
C & 0.6 & 0.95 &  0.010 & 1.000 &  17.830 & 0.304 & 2.869 & 0.973 & 0.216 & 4.476 & 1.685 & 1.448 & 3.853 & 1.760 & 1.483 & HYP\\
& 0.7 & 0.99 &  0.010 & 1.000 &  16.830 & 0.291 & 3.353 & 0.949 & 0.214 & 4.168 & 1.569 & 1.349 & 3.597 & 1.643 & 1.384 & HYP\\
& 0.8 & 1.04 &  0.010 & 1.000 &  15.920 & 0.279 & 3.918 & 0.922 & 0.213 & 3.926 & 1.477 & 1.270 & 3.391 & 1.549 & 1.305 & HYP\\
& 0.9 & 1.10 &  0.010 & 1.000 &  15.110 & 0.265 & 4.572 & 0.894 & 0.213 & 3.732 & 1.405 & 1.208 & 3.223 & 1.472 & 1.240 & HYP\\
& 1.0 & 1.16 &  0.010 & 1.000 &  14.440 & 0.252 & 5.285 & 0.866 & 0.214 & 3.576 & 1.346 & 1.157 & 3.087 & 1.410 & 1.188 & HYP\\
& 1.5 & 1.48 &  0.010 & 1.000 &  12.340 & 0.190 & 9.915 & 0.739 & 0.217 & 3.124 & 1.176 & 1.011 & 2.683 & 1.226 & 1.032 & HYP\\ \hline
\end{tabular}}
\end{table*}
 
We search for the maximum rest mass models for
$\hat{A}^{-1}\in[0.0,1.0]$ in increments of 0.1, as well as for
$\hat{A}^{-1} = 1.5$. We also build the benchmark TOV limit model
($M^{TOV}_{0, \rm max}$), and the supramassive limit model ($M^{sup}_{0,\rm max}$) against which we compare the increase in rest mass
when considering differential rotation. For these same models, we also
consider the increase in the gravitational mass compared to the
gravitational mass of the TOV limit ($M^{TOV}_{\rm ADM, max}$) and the
supramassive limit ($M^{sup}_{\rm ADM, max}$). As a reminder, the
values for $M^{TOV}_{0, \rm max}$, $M^{sup}_{0, \rm max}$,
$M^{TOV}_{\rm ADM, max}$, and $M^{sup}_{\rm ADM, max}$ for each
EOS in our sample are shown in Tab.~\ref{tab:stiffness}. To find the
maximum rest mass Type A and C models presented here we built sequences of constant
$\hat{A}^{-1}$ and $\epsilon_{\rm max}$ while varying
$\frac{r_p}{r_e}$ from 1.0 to 0.01 and
found the model with the largest rest mass. To find the maximum rest mass Type B models presented here we first built models at $\hat{A}^{-1} = 1.5$ and $\frac{r_p}{r_e} = 0.01$ (Type C models), then decreased $\hat{A}^{-1}$ to the target value, and finally increased $\frac{r_p}{r_e}$ to as high as possible. For each model type we then change the value
of $\epsilon_{\rm max}$ while holding $\hat{A}^{-1}$ fixed, and repeat
the aforementioned scans, resulting in a set of maximum rest mass
models for each value of $\epsilon_{\rm max}$ at a given value of
$\hat{A}^{-1}$. The model with the largest rest mass among these is
taken to be the maximum rest mass model for a given value of
$\hat{A}^{-1}$ and of a given type (A, B or C). We note that since we
are not able to build complete $\hat{A}^{-1}$-constant sequences for
Type B stars, the values we report for the Type B stars correspond to
the maximum rest mass configurations found in our search.

Properties of the maximum rest mass models are shown in
Tabs.~\ref{tab:maxmass_FPS}-\ref{tab:maxmass_APR}. We find that for
the four EOSs considered here, the maximum rest mass model is the
configuration with $\hat{A}^{-1} = 0.4$ (the lowest value of
$\hat{A}^{-1}$ for which Type B models exist that we considered) and
is always a Type B model. For polytropes,
~\cite{Gondek-Rosinska:2016tzy} showed that Type B models at the
lowest possible value of $\hat{A}^{-1}$ are the most massive ones,
too.  As shown in Tabs.~\ref{tab:maxmass_FPS}-\ref{tab:maxmass_APR},
depending on the EOS \"ubermassive configurations arise not only for Type B, but also for Type C stars. Our search results suggest that \"UMNSs are, in general, more common for softer EOSs, which is consistent with our finding that softer EOSs lead to larger increases in the rest mass.

\begin{figure}[ht]
   \includegraphics[width=8.5cm]{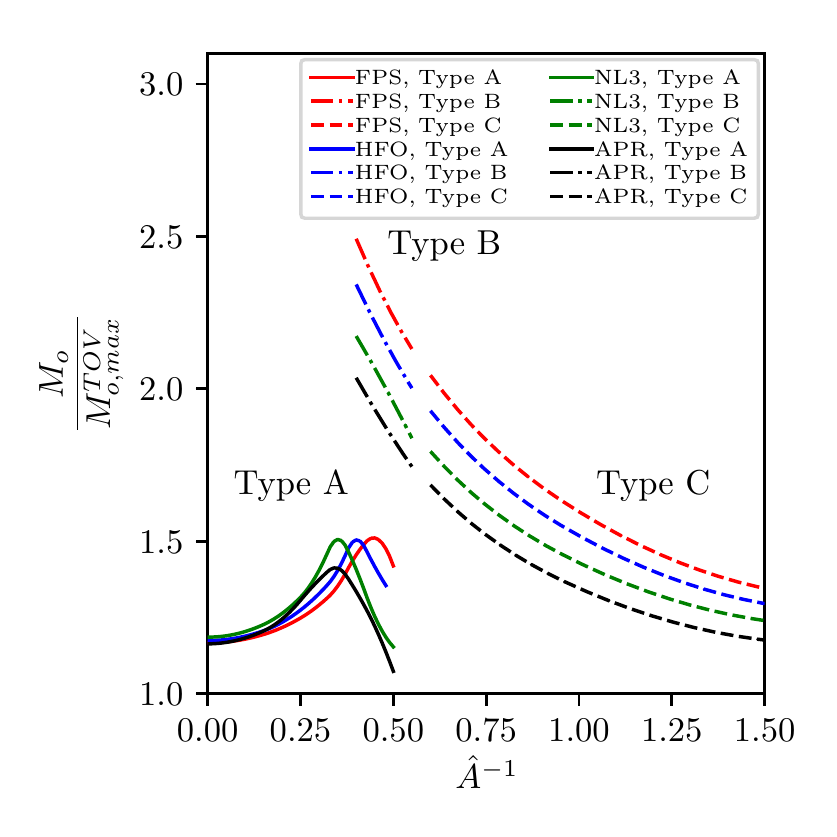}
   \caption{ Ratio of rest mass $M_0$ to maximum TOV rest mass
     $M_{0, \rm max}^{TOV}$ as a function of degree of differential
     rotation $\hat{A}^{-1}$ for each solution type and EOS. The solid
     lines show the relative increase for Type A models, the
     dash-dotted lines show the relative increase for Type B models,
     and the dashed lines show the relative increase for Type C
     models. The red, blue, green, and black lines correspond to the FPS, HFO, NL3, and APR EOSs, respectively. }
   \label{fig:dMo}
 \end{figure}

We now compare our results for the APR and FPS EOSs with those of
\cite{Morrison:2004fp}. In~\cite{Morrison:2004fp} models of
differentially rotating stars were constructed that exceeded the TOV
limit rest mass by at most $31\%$ for APR and $46\%$ for
FPS. Given that the maximum rest mass models for APR and FPS reported
  in~\cite{Morrison:2004fp} correspond to $\hat A^{-1}=0.3$, and $\hat
  A^{-1}=0.5$, respectively, it suggests that these models were of
  Type A. However, maximum rest mass Type B models (quasi-toroidal
models at low degree of differential rotation) are in all cases more
massive than maximum rest mass Type A and C models. When considering
Type B models, we find that the maximum rest mass can increase by as
much as approximately $100\%$ and $150\%$ in the cases of APR and FPS,
respectively. The largest increase in rest mass for NL3 and HFO is
approximately $120\%$ and $130\%$, respectively. These maximum rest mass
configurations are all \"UMNSs. We emphasize the fact
that generally Type B models tend to be the most massive, and that
they show the largest increase in rest mass when compared to the TOV
limit, as depicted in Figure~\ref{fig:dMo}.

We find that among Type A models, those with larger values of
$\hat{A}^{-1}$ tend to have larger rest mass. However,
the relationship between $\hat{A}^{-1}$ and $M_0$ for Type A models is
not monotonic. There appears to be a value of $\hat{A}^{-1}$
above which the maximum rest mass begins to decrease as seen from the curves in the lower left corner of Fig.~\ref{fig:dMo}. This feature of the solution
space was also observed in \cite{Studzinska:2016ofb} for a $\Gamma =
2.5$ polytrope, suggesting that it may arise for stiffer EOSs. We note
that this feature is observed for all EOSs we study here,
which have effective polytropic exponents of $\Gamma_{eff}^{nuc}
\gtrsim 2.5$. For Type A the largest rest mass models were found for a
value of $\hat{A}^{-1}$ of 0.3 for APR, 0.35 for NL3, 0.4 for HFO, and 0.45 for FPS, respectively, suggesting that the value of $\hat{A}^{-1}$ at which the maximum rest mass begins to decrease is smaller for stiffer EOSs (Note that we also built maximum rest mass Type A models in increments of $\hat{A}^{-1} = 0.05$ for finer resolution in Fig. ~\ref{fig:dMo}).
 
For Type B and C models, we observe the same monotonic behavior between the
increase in rest mass relative to the TOV limit and $\hat A^{-1}$ as
seen for stiffer EOS in \citep{Studzinska:2016ofb}, i.e., the
  maximum rest mass increases with decreasing $\hat A^{-1}$. We also
find the same general ordering of EOS by stiffness whereby softer EOSs (FPS and HFO)
tend to exhibit larger increases of the rest mass compared to the TOV
limit (see Fig.~\ref{fig:dMo}). 

It is noteworthy that the largest increase in rest mass is seen in the
FPS EOS, the softest EOS in our set. In \cite{Studzinska:2016ofb} the
maximal increase in rest mass was observed for a moderately stiff EOS
which was neither the softest nor the stiffest considered. There it
was argued that, generally, the increase in rest mass compared to the
TOV limit due to differential rotation decreases with increasing
stiffness. We observe the same trend with realistic EOSs for Type B
and C models, which indicates that the largest increase in rest mass
due to differential rotation is possible for quasi-toroidal
configurations described by softer EOSs. 
For the polytropes considered in~\cite{Studzinska:2016ofb}, it was
found that stiffer EOSs show larger increases in the rest mass for
Type A models of \textit{low} $\hat{A}^{-1}$ (0.0 to 0.4)
We find a similar general trend for the Type
A models at low $\hat{A}^{-1}$ (0.0 to 0.3) presented here. However, an ``anomaly'' in
this trend is seen in the case of the APR EOS. For instance, the low $\hat{A}^{-1}$ (0.0 to 0.3) maximum rest mass models for the FPS
EOS (the softest considered here) show a very similar increase in the
rest mass as those of APR (the stiffest EOS considered here), as can
be seen from the low $\hat{A}^{-1}$ part of the leftmost curves of
Figure~\ref{fig:dMo} and the corresponding $\dfrac{M_0}{M_{0, \rm max}^{TOV}}$ entries of Tables \ref{tab:maxmass_FPS} and \ref{tab:maxmass_APR} for Type A models of low $\hat{A}^{-1}$. A possible explanation for the break in the
trend is that \cite{Studzinska:2016ofb} consider a large range of
adiabatic indices $1.8 \leq \Gamma \leq 3.0$, whereas the effective
adiabatic indices of the EOSs in our sample cover a smaller range.  On
the other hand, assigning one number to stiffness in the case of
realistic EOSs may not be entirely appropriate as the stiffness
defined through stellar models may depend on the choice of
mass. For example, the TOV mass-radius curves of the HFO and APR EOSs
intersect near their corresponding TOV limits (see
Appendix~\ref{app:mrcurves}).  In this work we defined stiffness based
on the maximum rest mass TOV configurations and on TOV configurations with
gravitational mass of 1.4$M_\odot$. It is also the case that Type A
configurations of low $\hat{A}^{-1}$ mostly sample the value of
$\epsilon_{\rm max}$ from higher density regions of the EOSs which may
be of comparable stiffness. This is supported by the fact that in all
cases considered here, the values of $\epsilon_{\rm max}$ for the Type
A models of low $\hat{A}^{-1}$ are larger than for the Type B and C
models. The anomaly we mentioned above would not be observed for EOSs of constant
stiffness as defined by the effective polytropic exponent, as in the
case of the polytropes of fixed polytropic index studied
in~\cite{Studzinska:2016ofb}. A systematic study of the effect may
employ realistic EOSs as done here or a piecewise polytropic EOS such
as those presented in \cite{Read2009, Ozel:2009da, Raithel:2016bux}, where the polytropic
index has a dependence on the energy density. However, such a study goes beyond the scope
of the current work.

\section{Conclusions and discussion}
\label{sec:conclusions}

In this paper, we have presented results for the solution space of
general relativistic differentially rotating neutron stars with
realistic EOSs. We found that the different types of differentially
rotating equilibrium solutions that were previously discovered for
polytropes~\cite{Gondek-Rosinska:2016tzy,Studzinska:2016ofb} with the
KEH rotation law~\cite{KEH1989MNRAS.237..355K}, exist for realistic
neutron star equations of state, too. Moreover, we demonstrated that
codes based on the KEH scheme~\cite{KEH1989MNRAS.237..355K}, such as
the Cook code~\cite{CST94a,CST94b}, can build these different types of
stars, although we were not able to construct Type D sequences of
constant degree of differential rotation and constant maximum energy
density or complete Type B sequences. The Cook code is capable of
building most of the extremely massive quasi-toroidal, relativistic
configurations using realistic EOSs, but finds it challenging to converge on
solutions which are both highly pinched and quasi-toroidal. Note that Type
D stars are not likely to be
physical~\cite{Gondek-Rosinska:2016tzy,Studzinska:2016ofb}. 

We presented the maximum rest mass configurations found in our search of
the solution space for three of the four types of solutions we were
able to construct. As
in~\cite{Gondek-Rosinska:2016tzy,Studzinska:2016ofb} we find
configurations that can support a mass more than 2 times the TOV limit. We
called these configurations ``\"ubermassive''. For the equations of
state considered here we find that \"ubermassive stars can support up
to 150\% more rest mass than the TOV limit mass with the same equation
of state. This number is a lower limit to the maximum rest mass that can be
supported by differential rotation. We have classified the maximum
mass configurations we found as supramassive, hypermassive or \"ubermassive, and
  found that depending on the equation of state \"ubermassive stars
  can be Type B or Type C.

Differentially rotating hypermassive neutron stars can form following
binary neutron star mergers. Clearly, following such a merger, the
remnant configuration cannot have mass more than 2 times the TOV limit
mass. Thus, the \"ubermassive configurations we found may never appear
in Nature, and if they do they would have to form through some more
exotic channel. Moreover, it is well known that in binary neutron star
mergers there exists a threshold value for the binary total mass above
which a black hole forms promptly after
merger~\cite{ST,STU1,STU2,Hotokezaka:2011dh,Bauswein:2013jpa,Bauswein:2017aur}.
This value for the treshold mass ($M_{\rm thres}$) depends on the
equation of state, and for quasicircular, irrotational binaries it may
be up to $\sim 70\%$ greater than the TOV limit mass~\cite{ST}. It may
also be that for irrotational binaries $M_{\rm thres} \in
[2.75-3.25]M_\odot$~\cite{Paschalidis:2018tsa}. Therefore, it may be
difficult to form even extreme hypermassive neutron stars in binary
neutron star mergers. An exception may be dynamical capture mergers
such as those studied recently
in~\cite{Gold:2011df,East:2012ww,Paschalidis:2015mla,PEFS2016,East:2016zvv,Radice:2016dwd,Chaurasia:2018zhg,Papenfort:2018bjk},
where the total angular momentum at merger can be higher than those in
quasicircular binaries, which can provide additional centrifugal
support.

Regardless of the precise value of $M_{\rm thres}$ the question about
what type of differentially rotating star can form following a neutron
star merger remains open. This is interesting because less dramatic,
but significant increases to the maximum supportable mass can arise
for degrees of differential rotation different than those
corresponding to the more extreme cases. Such configurations may be
relevant for binary neutron star mergers, and may have implications
for the stability and lifetime of their hypermassive neutron star
remnants.

Another important question is how well the KEH rotation law describes
the differential rotation profile of a hypermassive neutron star
formed in a binary neutron star merger and whether the different types
of stellar solutions are unique to the KEH law. The rotational
properties of hypermassive neutron stars formed in quasicircular
binary neutron star mergers have been studied recently in a number of
works~\cite{Kastaun:2014fna,Kastaun:2016yaf,Kastaun:2016yaf,Kastaun:2016elu,Hanauske:2016gia,Ciolfi:2017uak}
and they appear to deviate from that of the KEH rotation
law. Nevertheless, the rotation profiles reported
in~\cite{Paschalidis:2015mla,PEFS2016} for eccentric neutron star
mergers are different and seem to be within the realm of the KEH
rotation law. Interestingly, the remnants found
in~\cite{Paschalidis:2015mla,PEFS2016} were also quasi-toroidal. In a
recent work a new differential rotation law was
introduced~\cite{Uryu:2017obi} which captures the rotational profile
of some binary neutron star merger remnants. An interesting follow up
to our work is to adopt this new rotation law and investigate the
maximum possible mass that can be supported for different realistic
EOSs and whether different types (or even more types) of
differentially rotating stars arise.

Finally, the issue of dynamical stability of the different types of
differentially rotating stars is important to address. Moreover, are
\"ubermassive stars dynamically stable? Many of the equilibrium
configurations we built have $T/|W| > 0.25$, and hence are unstable to
a dynamical bar mode instability (see ~\cite{Paschalidis:2016vmz} and
references therein). Some of the configurations we built have
dimensionless spin parameter $J/M^2 > 1$, which does not necessarily
imply collapse on a secular timescale, as the star can be unstable to
non-axisymmetric modes and collapse through fragmentation (see
~\cite{Zink:2006qa,Reisswig:2013sqa} and~\cite{Paschalidis:2016vmz} for a review). Non-axisymmetric
instabilities in differentially rotating stars arise even for low
values of
$T/|W|$~\cite{Centrella2001,Saijo2003,Watts2005,Saijo2006,Ou2006,Corvino2010,Saijo:2016vcd,Yoshida:2016kol,Saijo:2018thy}
and in binary neutron star merger
remnants~\cite{Paschalidis:2015mla,PEFS2016,East:2016zvv,Radice:2016gym,Lehner:2016wjg}.
If a certain type of solution is dynamically unstable to collapse,
then it cannot arise in Nature, despite the fact that the equilibrium
configuration can support an amount of mass much larger than the TOV
limit. Unlike the case of uniformly rotating stars the turning point
theorem~\cite{1981ApJ...249..254S,1988ApJ...325..722F,Schiffrin:2013zta}
does not apply to differentially rotating stars (although it seems to
apply approximately for type A
configurations~\cite{Weih:2017mcw,Bozzola:2017qbu}), therefore
dynamical simulations in full general relativity offer a
straightforward avenue to study the dynamical stability of these
configurations. The solutions we have constructed can serve as initial
data for such dynamical simulations. We will address all of these open
questions in future studies.\\

\section*{Acknowledgments}

We are grateful to Stuart L. Shapiro for access to the equilibrium
rotating NS code. We thank Gabriele Bozzola for useful discussions,
and Nick Stergioulas for access to his RNS code \cite{Stergioulas:1994ea, Bauswein:2017aur} which we used in an
attempt to build some more extreme configurations. Calculations were
in part performed on the Ocelote cluster at the University of Arizona,
Tucson. This work made use of the Extreme Science and Engineering
Discovery Environment (XSEDE), which is supported by National Science
Foundation, through grant number TG-PHY180036.

\appendix
\section{Calculation of polytropic representation of realistic Euations of State}\label{sec:appendixA}

When building polytropic stellar configurations in geometrized units,
the polytropic constant $\kappa$ defines a fundamental length scale
($\kappa^{n/2}$) which scales out of the problem. To calculate
$\kappa_{eff}^{nuc}$ we first build the maximum rest mass TOV models for
polytropes with $\Gamma_{eff}^{nuc}$ as defined in Equation
\eqref{eq:Gamma_nuc}. Next, we calculate the polytropic constant in
geometrized units $\kappa_{eff, geo}^{nuc}$ by matching the maximum
TOV ADM masses of the nuclear and polytropic EOSs,
\begin{equation}\label{eq:kappa_geo}
\kappa_{eff,geo}^{nuc} = \left( \dfrac{M_{\rm ADM,max}^{TOV,nuc}}{M_{\rm ADM,max}^{TOV,poly}} \right)^{\dfrac{2}{n_{eff}^{nuc}}}.
\end{equation}

The quantity in the parentheses of Equation ~\eqref{eq:kappa_geo} 
is then converted to a unit of length (specifically, we work in cgs units). 
We then replace the appropriate factors of $G$ and $c$ needed to express 
our physical quantities in cgs units,
\begin{equation}\label{eq:kappa_cgs}
\kappa_{eff}^{nuc} = \dfrac{G^{\frac{1}{n}}}{c^{\frac{2}{n} - 2}} \kappa^{nuc}_{eff, geo}.
\end{equation}

Finally, we write the polytropic representation of the nuclear 
EOSs we considered as

\begin{equation}\label{eq:polyeff_p}
P = \kappa_{eff}^{nuc} \rho_0^{\Gamma_{eff}^{nuc}}
\end{equation}
and

\begin{equation}\label{eq:polyeff_e}
\epsilon = \rho_0c^2 + \dfrac{P}{(\Gamma_{eff}^{nuc} - 1)},
\end{equation}
where $\Gamma_{eff}^{nuc}$ is the effective adiabatic index as calculated in Section \ref{sec:EOSs}.
\section{Mass-radius curves for realistic equations of state}
\label{app:mrcurves}
\begin{figure}\label{fig:MR}
\includegraphics[width=8.5cm]{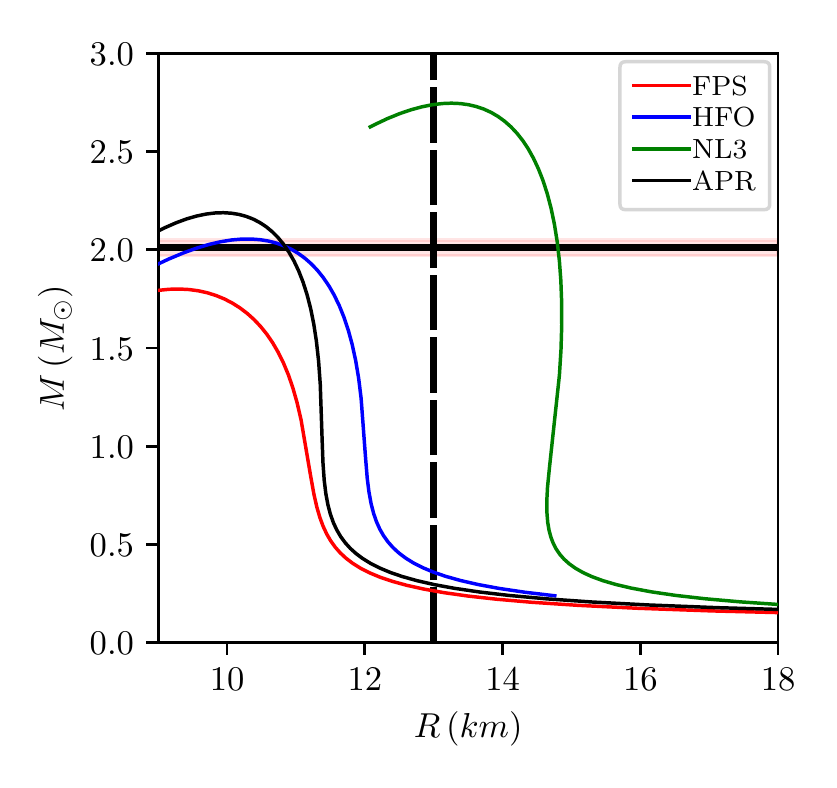}\caption{Mass-radius relation for the realistic EOSs used in this work. The red, blue,
  green, and black lines corresponds to the FPS, HFO, NL3, and APR
  EOSs, respectively. The solid horizontal line and red horizontal
  band correspond to the upper bound on the NS mass from observations of PSR
  J1614-2230 \cite{Demorest2010}. The vertical solid line corresponds
  to upper limit on the NS radius from considerations of the tidal
  deformability as inferred from GW170817\cite{Raithel:2018ncd}.}
\end{figure}

Here we present the mass-radius relation of the nuclear equations of state used in
this work. As can be seen from Figure \ref{fig:MR}, all EOSs but the
FPS EOS respect the upper bound set on NS masses from observations of the
most massive pulsar to date, PSR J1614-2230 \cite{Demorest2010, Antoniadis2013}. Despite the FPS EOS having a maximum mass which falls below this upper bound we include it in
this study to offer a comparison to the results of
\cite{Morrison:2004fp}. It is also useful to consider the FPS EOS as an example of a relatively soft nuclear EOS. We find that the maximum increase in rest mass when compared to the TOV mass for the FPS EOS is the highest ($150\% $) in the set of EOS we considered (see Fig. \ref{fig:dMo}), which is consistent with our finding that softer EOSs result in larger increases of the rest mass relative to the TOV mass.

All EOSs but the NL3 EOS respect the 90\%
confidence upper bound on NS radii set by the tidal deformability of
NSs as inferred from GW170817 ~\cite{TheLIGOScientific:2017qsa,Abbott:2018exr,Raithel:2018ncd}. Despite the fact that using the NL3 EOS results in stars with radii above this upper bound we include the it in this study to investigate the solution space of differentially rotating stars and maximum rest mass solutions for an EOS with a relatively large TOV mass. It is also useful to consider the NL3 EOS to investigate the solution space of differentially rotating stars for a relatively stiff EOS. We find that the maximum increase in rest mass when compared to the TOV mass for the NL3 EOS is among the lowest ($120\% $) in the set of EOS we considered, which is consistent with our finding that stiffer EOSs result in smaller increases of the rest mass relative to the TOV mass.

\bibliography{ref}

\end{document}